\documentclass{emulateapj}
\usepackage{graphicx, amssymb}

\newcommand{\Rsun}{\mbox{$\rm{R}_{\odot}$}}

\newcommand{\chisq}{\mbox{$\chi^{2}$}}

\slugcomment{Submitted to ApJ}
\shorttitle{White Dwarf Disk Lifetimes}
\shortauthors{Girven et al.}

\begin{document}

\title{Constraints on the Lifetimes of Disks Resulting from Tidally
Destroyed Rocky Planetary Bodies}


\author{J. Girven$^{1\star}$, C. S. Brinkworth$^{2,3}$, J. Farihi$^4$,
B. T. G\"ansicke$^1$, D. W. Hoard$^2$, T. R. Marsh$^1$, and D.
Koester$^5$}

\affiliation{$^{1}$ Department of Physics, University of Warwick,
Coventry CV4 7AL, UK}

\affiliation{$^{2}$ \textit{Spitzer} Science Center, California
Institute of Technology, Pasadena, CA 91125, US}

\affiliation{$^{3}$ NASA Exoplanet Science Institute, California
Institute of Technology, Pasadena, CA 91125, US}

\affiliation{$^{4}$ Department of Physics and Astronomy, University of
Leicester, Leicester LE1 7RH, UK}

\affiliation{$^{5}$ Institut f\"{u}r Theoretische Physik und
Astrophysik, University of Kiel, 24098 Kiel, Germany}

\email{$^{\star}$ j.m.girven@warwick.ac.uk}

\begin{abstract}

\textit{Spitzer} IRAC observations of 15 metal-polluted white dwarfs
reveal infrared excesses in the spectral energy distributions of
HE\,0110$-$5630, GD\,61, and HE\,1349$-$2305. All three of these stars
have helium-dominated atmospheres, and their infrared emissions are
consistent with warm dust produced by the tidal destruction of (minor)
planetary bodies.
This study brings the number of metal-polluted, helium and hydrogen
atmosphere white dwarfs surveyed with IRAC to 53 and 38 respectively.
It also nearly doubles the number of metal-polluted helium-rich white
dwarfs found to have closely orbiting dust by \textit{Spitzer}.
From the increased statistics for both atmospheric types with
circumstellar dust, we derive a typical disk lifetime of
$\log[t_\mathrm{disk}\mathrm{(yr)}]=5.6\pm1.1$ (ranging from $3\times10^4 -
5\times10^6$\,yr). This assumes a relatively constant rate of accretion
over the timescale where dust persists, which is uncertain. 
We find that the fraction of highly metal-polluted helium-rich white
dwarfs that have an infrared excess detected by \textit{Spitzer} is only
23\,per\,cent, compared to 48\,per\,cent for metal-polluted hydrogen-rich
white dwarfs, and we conclude from this difference that the typical
lifetime of dusty disks is somewhat shorter than the diffusion time scales
of helium-rich white dwarf. We also find evidence for higher
time-averaged accretion rates onto helium-rich stars compared to the
instantaneous accretion rates onto hydrogen-rich stars; this is an
indication that our picture of evolved star-planetary system interactions
is incomplete.
We discuss some speculative scenarios that can explain the observations.

\end{abstract}


\section{Introduction}
\label{s-int}

Over the past decade it has become increasingly clear that planetary
systems survive, at least in part, the late evolution of their host stars.
The observational evidence supporting this conclusion comes from metal
pollution observed in white dwarf atmospheres, and the commonly detected
circumstellar disks of solid and gaseous debris \citep[e.g.][Brinkworth
et al. 2012 submitted, and a host of other references
therein]{zuckerman+becklin87-1, kilicetal06-1, gaensickeetal06-3,
gaensickeetal07-1, gaensickeetal08-1, juraetal07-1, vonhippeletal07-1,
brinkworthetal09-1, farihietal10-1, debesetal11-1, farihietal11-1,
melisetal11-1}.

Due to the high surface gravities and limited radiative forces, heavy
elements sink rapidly within the atmospheres of cool ($T_{\rm eff} <
25,000$\,K) white dwarf stars, resulting in essentially pure hydrogen
or helium spectra \citep{koester09-1}. The downward diffusion of
metals occurs over a timescale that is typically only days to years
for hydrogen-rich (DA type) atmospheres (above $\sim11,000$\,K). Despite
this, a notable fraction of cool DA white dwarfs exhibit absorption
features due to the presence of trace metals \citep{zuckermanetal03-1,
koesteretal05-2}, and these must be the result of external sources and a
sign of ongoing accretion \citep{sionetal90-1}.

In numerous cases, the origin of these pollutants have been unambiguously
identified as circumstellar dust, primarily via \textit{Spitzer} studies
\citep[e.g.][]{reachetal05-1, juraetal09-1, farihietal09-1,
farihietal10-1}. Gaseous disks components are detected around a number of
stars, and the analysis of their dynamics demonstrates that the material
is located within $\sim 1 \Rsun$ \citep{gaensickeetal06-3,
gaensickeetal07-1, gaensickeetal08-1}, where large solid bodies should be
destroyed by tidal gravitational forces \citep{davidsson99-1}. The
chemical abundances of these photospheric pollutants can be analytically
linked to those of the accreted matter \citep{koester09-1}. In a number of
cases, the closely orbiting dust has been confirmed to be silicate-rich,
and typical of the material associated with planet formation \citep[e.g.
zodiacal and cometary dust;][]{reachetal05-1,juraetal09-1, reachetal09-1}.
Studying planetary systems around white dwarfs therefore unlocks the
potential to measure the bulk chemical composition of destroyed, and
subsequently accreted, rocky planetary bodies such as asteroids, moons, or
possibly major planets \citep[e.g.][]{zuckermanetal07-1, dufouretal10-1,
farihietal11-1, kleinetal11-1}.

The favored and successful model of a tidally destroyed asteroid
\citep{grahametal90-1, jura03-1} is consistent with the observed disk
properties \citep{gaensickeetal06-3, farihietal10-2, melisetal11-1,
debesetal11-1}, the subsequent photospheric pollution, and the
composition of both the orbiting and accreted material
\citep[e.g.][]{kleinetal11-1, zuckermanetal11-1}. Such a catastrophic
destruction is most readily achieved by a remnant planetary system
with at least one major planet \citep{debesetal02-1} that perturbs a
belt of smaller objects. Thus white dwarfs with disks and photospheric
metals may harbor complex planetary systems.

A significant uncertainty is the typical lifetime of the dust disks
\citep{jura08-1, kilicetal08-1}, yet this is an important indicator of
the mass of the parent body (or bodies) that generated the observed
debris. There are numerous stars with cooling ages greater than
1\,Gyr that exhibit atmospheric metal pollution, such as the
prototype, vMa\,2 \citep{vanmaanen17-1, greenstein56-1,weidemann60-1}.
\citet{koesteretal11-1} recently identified 26 white dwarfs with
temperatures $5000-8000$\,K and $10^{20} - 10^{23}$\,g of accreted
metals in their convection zones. These stars have typical
($\log\,g=8$) cooling ages of $1-6$\,Gyr but must have accreted
material recently, within the past few million years. Interestingly,
there is only a single (and anomalous) infrared excess around stars
with cooling ages older than 1\,Gyr
\citep[G166-58;][]{farihietal08-1}, so this may be viewed as an upper
limit for typical disk lifetimes.

From theoretical considerations, \citet{rafikov11-1} finds the
lifetimes of the compact dust disks around white dwarfs should be of
the order $10^6$\,yr, when dominated by Poynting-Robertson
drag. However, Poynting-Robertson drag cannot produce the highest
(average) accretion rates inferred for the helium atmosphere stars (DB)
with metals, which are on the order of $10^{10}-10^{11}$\,g\,s$^{-1}$
\citep{farihietal10-1}. \citet{xu+jura12-1} extend the estimates of
\citet{rafikov11-1} and find a factor of five higher accretion rates, but
this still does not lead to high enough rates. To produce higher accretion
rates, gas resulting from sublimated dust, produced at the inner edge of
the dust disk \citep{jura08-1, farihietal09-1}, efficiently transports
angular momentum outward and fuels a more rapid in-fall of material.
\citet{rafikov11-2} calculates that the lifetime of a $10^{22}$\,g disk is
reduced to several $10^4$\,yr.

We report the results of a combined \textit{Spitzer} IRAC Cycle 5 and 6
search for disks around 15 cool, metal-contaminated white dwarfs. Three
of the observed stars show infrared excess that is the trademark of warm
dust orbiting within the tidal disruption radius: HE\,0110$-$5630, GD\,61,
and HE\,1349$-$2305. These three stars significantly increase the number
of metal-polluted DB white dwarfs with disks and enable a statistical
estimate of a typical disk lifetime based on comparison with their DA
counterparts. The observations are described in \S\,\ref{s-obs}, the
analysis and disk modeling in \S\,\ref{s-res}. In \S\,\ref{s-dis} the
statistics of dusty white dwarfs is revisited, and our method for
estimating disk lifetimes is described.

\section{Observations}
\label{s-obs}

We were awarded \textit{Spitzer Space Telescope} time in Cycle 5
(program 50340) and Cycle 6 (program 60119), to search for infrared
excesses from circumstellar dust around 16 white dwarfs published as
metal-enriched (Table\,\ref{t-tar}). These targets were chosen from
the literature, and significantly increase the number of DBZ (for
simplicity, we use `DBZ' to refer to all helium-rich subtypes with
metals, including those with trace hydrogen) stars observed with
\textit{Spitzer} IRAC, bringing their numbers onto par with those of
the DAZ stars. One of the targets, PHL\,131, was subsequently found by
one of us (DK) to be a $40,000$\,K white dwarf with interstellar
calcium absorption, and thus we report its fluxes, but exclude it from
the rest of the study.

Imaging observations were obtained for each of the white dwarf targets
using the Infrared Array Camera \citep[IRAC;][]{fazioetal04-1}. Exposures
were taken in the 4.5 and 7.9\,$\mu$m channels using the medium-$ $scale,
cycling dither pattern for the Cycle 5 targets objects and in the 3.6 and
4.5\,$\mu$m channels for the Cycle 6 objects. For GD\,61 and NLTT\,51844,
images were also obtained with the blue Peak-Up Imaging array of the
Infrared Spectrograph at $15.6\,\mu$m, this time using the small-scale,
cycling dither pattern.

Three of the target white dwarfs were chosen from the Hamburg Schmidt
\citep[HS;][]{hagenetal95-1} and the Hamburg European Southern Observatory
Schmidt \citep[HE;][]{wisotzkietal96-1} quasar surveys. As discussed in
\citet{farihietal10-1}, the SIMBAD coordinates for these objects are often
inaccurate by up to a few arc minutes, and Table\,\ref{t-hshe} gives
correct positions for these sources based on their IRAC images.

\subsection{Data Analysis}
\label{ss-red}

The IRAC image photometry was performed on the individual BCD frames
downloaded from the \textit{Spitzer} archive and reduced using S18.7.0. 
These were corrected for array-location dependence, as described in the
IRAC Data Handbook\footnote{See http://ssc.spitzer.caltech.edu/irac/dh}. 
We used the point source extraction package \textsc{apex multiframe}
within MOPEX \citep{makovozetal06-1} to perform PSF fitting photometry on
the individual BCDs. We do not apply a color correction because we use the
isophotal wavelengths in our analysis. The magnitude of the correction is
therefore negligible compared to our uncertainties. The pixel
phase correction for channel\,1 data is also minimal in comparison to the
uncertainties and therefore was not applied. A minimum 5\,per\,cent
uncertainty on the flux density was assigned to account for the systematic
uncertainties in this method \citep{reachetal05-1}.

\begin{figure}
\begin{center}
\includegraphics[width=0.4\columnwidth]{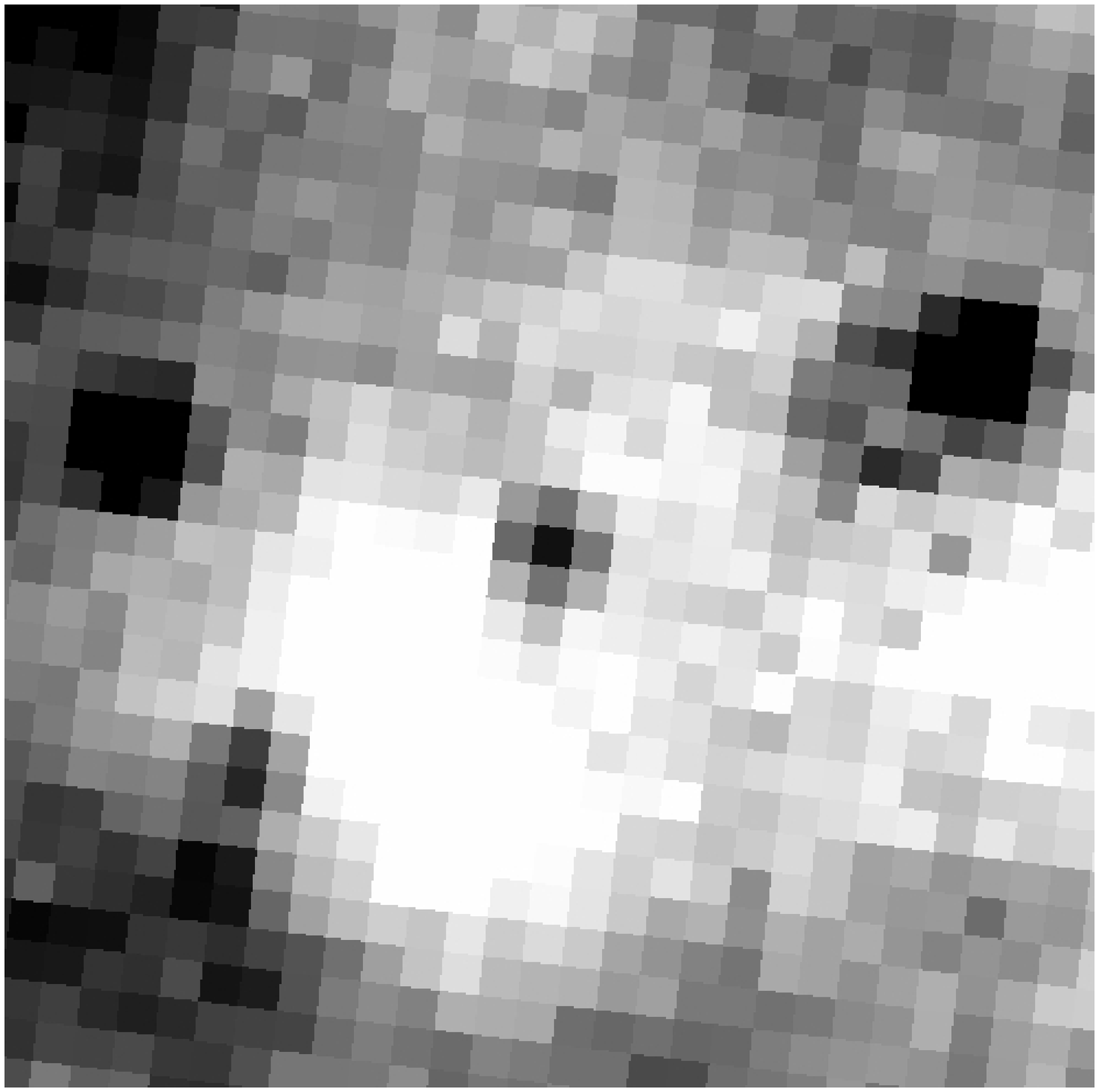}
\includegraphics[width=0.4\columnwidth]{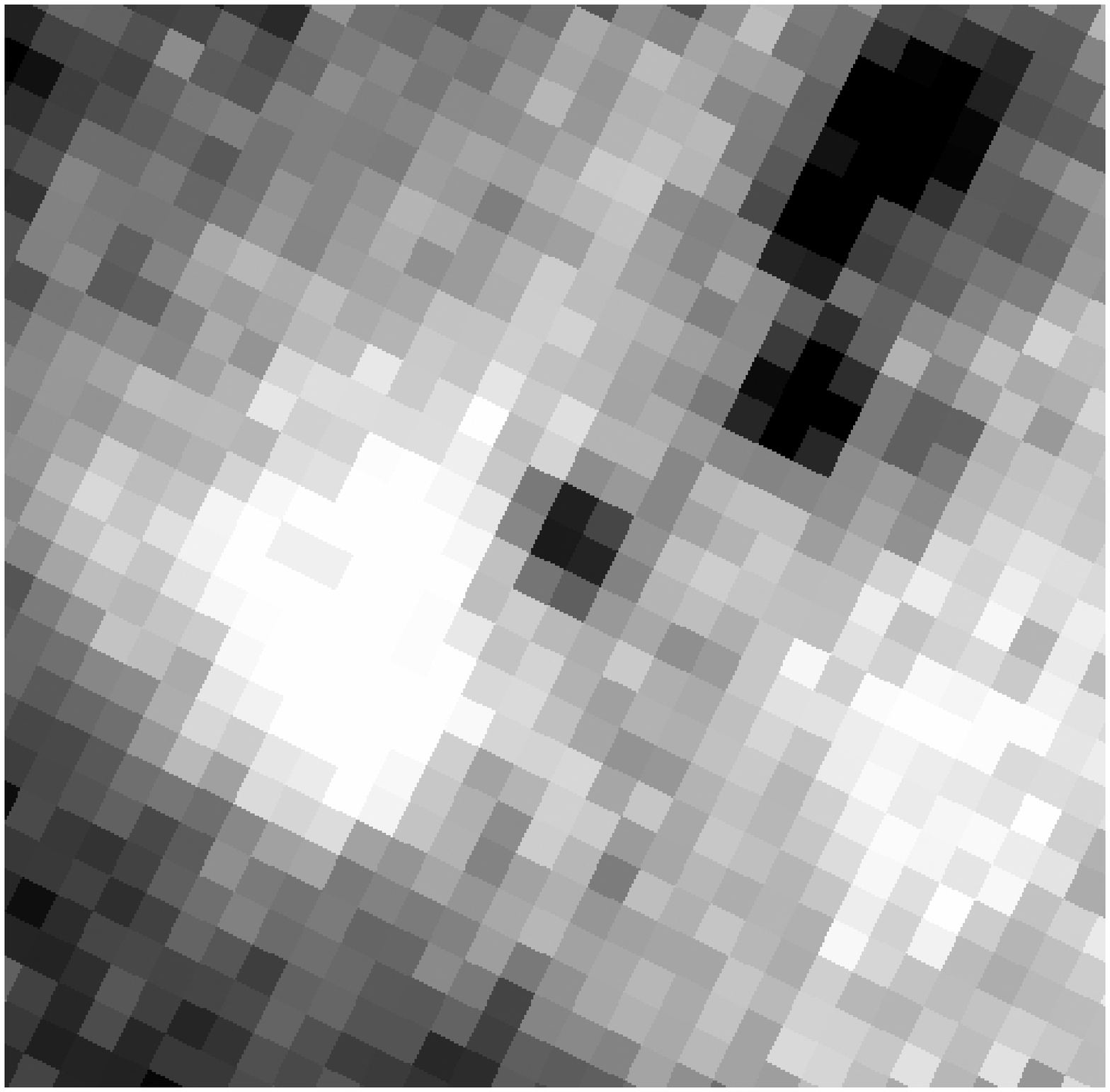}
\end{center}
\caption{\label{f-irs} IRS Peak-Up image mosaics of GD\,61 (\textit{left})
and NLTT\,51844 (\textit{right}). The images are orientated North up and
East left at $1\farcs8$ pixel$^{-1}$ and $50\arcsec$ across. The
background is clearly variable across the array and is therefore the
primary source of error (\S\,\ref{ss-red}).}
\end{figure}

The Infrared Spectrometer \citep[IRS][]{houcketal04-1} Peak-up Imaging was
analyzed according to the method of Brinkworth et al.\ (2011, submitted).
The sky background was found to be variable over the array (see
Figure\,\ref{f-irs}), so the individual BCDs were initially
median-combined to make both a flat and sky, using $3\sigma$ rejection.
This was then scaled to the overall median of the Post-BCD mosaic and
subtracted from each of the individual BCDs. To completely reduce the sky
level to zero, the edges of the image were temporarily discarded and a
median of the center of the image was found and removed. The sky can
therefore assumed to be zero and an infinite sky annulus is effectively
used during photometry. The flat-fielded and median-subtracted BCDs were
mosaicked using MOPEX with a dual-outlier rejection. The pixel scale for
the mosaic was set to the default $1\farcs8$ pixel$^{-1}$. The photometry
was performed with \textsc{apphot} within IRAF using a 3 pixel aperture
radius, but no sky subtraction. The aperture correction to the calibration
aperture sizes, as provided by the \textit{Spitzer} Science Center, was
performed and the fluxes were converted from MJy sr$^{-1}$ to mJy. As an
estimate of the error, a series of other apertures were placed around the
target on the mosaic and the standard deviation of fluxes was taken as the
uncertainty. All measured \textit{Spitzer} fluxes are listed in
Table\,\ref{t-flu}.

\subsection{Near-Infrared Observations}
\label{ss-ind}

Supplemental near-infrared photometry for most target stars was obtained
on the 23 March 2011 with the William Herschel Telescope using the
Long-Slit Intermediate Resolution Infrared Spectrograph (LIRIS;
\citealt{manchadoetal98-1}), and on 10$-$12 August 2011 with the New
Technology Telescope using Son of Isaac (SOFI;
\citealt{moorwoodetal98-1}). Images taken in a 9-point dither pattern were
obtained the $J$, $H$, and $K_s$-band filters with typical total exposure
times of 270\,s in clear conditions; 3 standard star fields were observed
in a similar manner for photometric zero-point calibration. These data
were reduced in the standard manner, by subtracting a median sky from each
image in the dithered stack, flat fielding (using sky flats), then
averaging and recombining frames.

LIRIS suffers from what is known as a detector reset anomaly, which
appears in certain frames as a discontinuous jump (in dark current)
between the upper and the lower two quadrants. To remove this unwanted
signal, after flat fielding and sky subtraction, the detector rows were
collapsed into a median column (with real sources rejected), and
subsequently subtracted from the entire two dimensional image. The
resulting fully-reduced frames exhibit smooth backgrounds, free of the
anomalous gradient. Analogously, SOFI exhibits significant cross-talk
between quadrants, which becomes apparent for bright sources. The IRAF
script \textsc{crosstalk} was employed on each of the raw frames prior to
processing as described above, effectively removing any unwanted
artifacts.

Aperture photometry of standard stars and relatively bright targets was
performed using $r\approx4\arcsec$ aperture radii and sky annuli of
ranging between $5\arcsec$ and $8\arcsec$. For relatively faint targets or
those with neighboring sources, smaller apertures were employed with
corrections derived from several brighter stars within the same image
field and filter. In a few cases, PSF-fitting photometry (i.e.,
\textsc{daophot}) was used as a second method in addition to photometry
with small apertures. All data taken in the $K_s$-band filter were
flux-calibrated using ARNICA \citep{huntetal98-1} $K$-band standard star
photometry. The measured photometry can be found in Table\,\ref{t-ind}.

\section{Analysis and Results}
\label{s-res}

\subsection{Removal of Nearby Background Source Flux}
\label{ss-bg}

\begin{figure}
\includegraphics[width=\columnwidth]{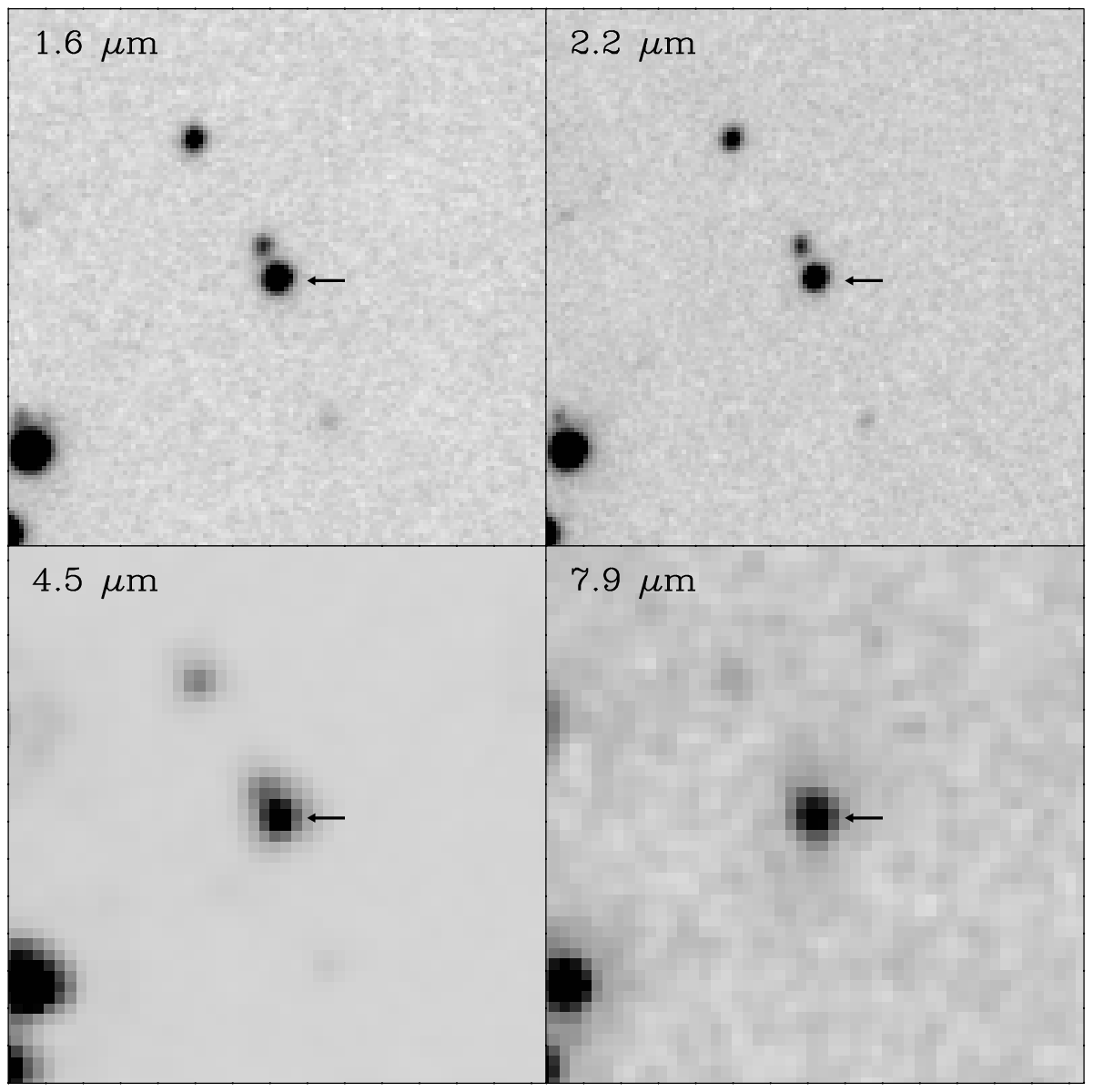}
\caption{\label{f-0435im} Infrared images of GD\,61 taken with LIRIS at
$H$ and $K_s$, and \textit{Spitzer} IRAC at 4.5 and $7.9\,\mu$m. All
images are oriented North up and East left, and are $30\arcsec$ across,
with $0\farcs25$ pixel$^{-1}$ for LIRIS and $0\farcs6$ pixel$^{-1}$ for
IRAC. The white dwarf (marked by the arrow) is separated from a
neighboring source by $1\farcs9$. The neighbor does not influence the
flux measurements presented in Table\,\ref{t-flu} (\S\,\ref{ss-bg}). The
neighboring source must be in the background as its brightness never
exceeds that of the white dwarf.}
\end{figure}

\begin{figure}
\begin{center}
\includegraphics[width=\columnwidth]{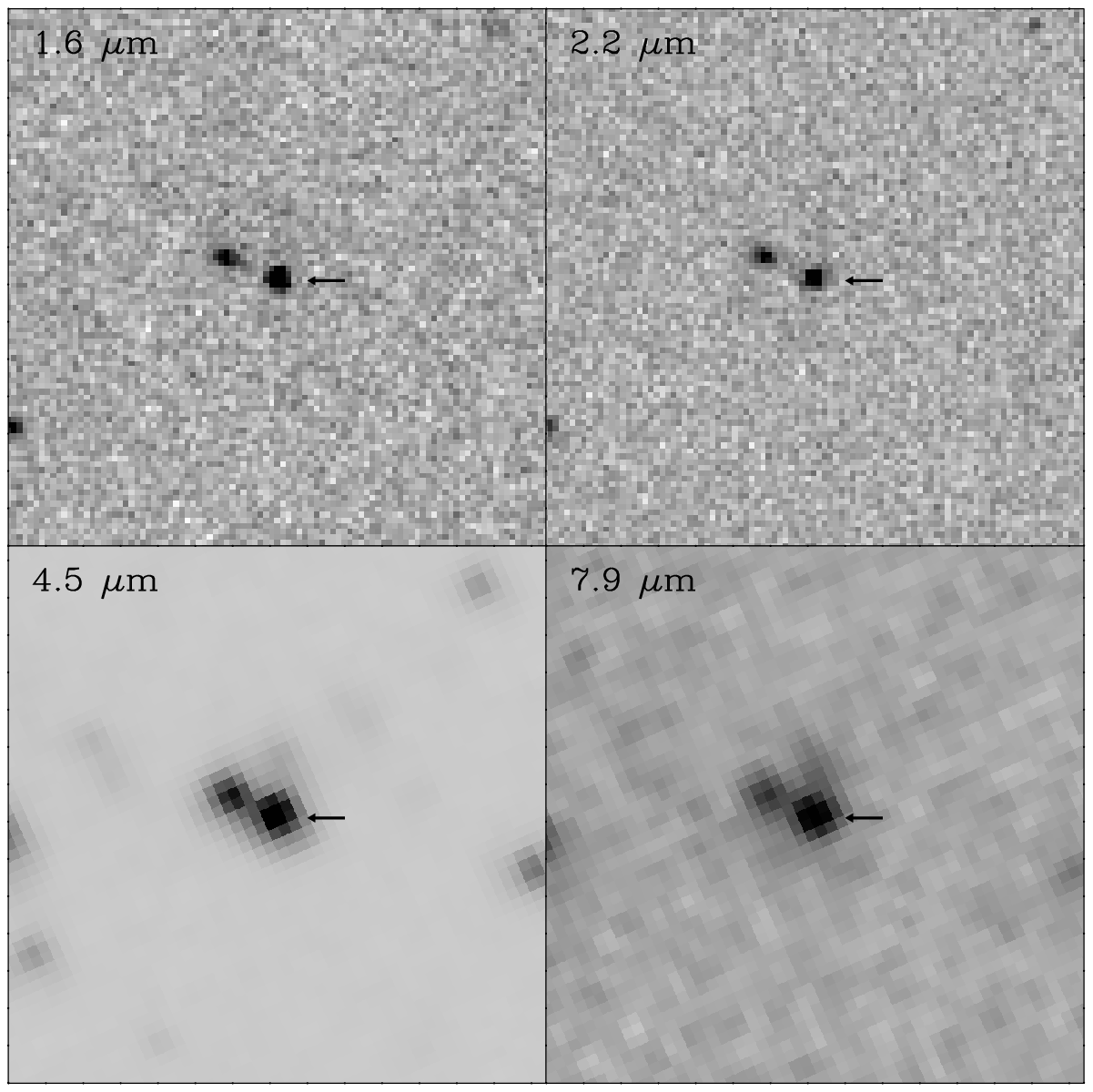}
\end{center}
\caption{\label{f-1349im} Infrared images of HE\,1349$-$2305 taken with
SOFI at $H$ and $K$, and \textit{Spitzer} IRAC at 4.5 and $7.9\,\mu$m. 
All images are oriented North up and East left, and are $30\arcsec$
across, with $0\farcs25$ pixel$^{-1}$ for LIRIS and $0\farcs6$
pixel$^{-1}$ for IRAC. The white dwarf (marked by the arrow) is separated
from a neighboring source by $2\farcs8$, but does not influence the flux
measurements presented in Table\,\ref{t-flu} (\S\,\ref{ss-bg}). The
neighbor is clearly extended in the $H$-band image and is therefore
extragalactic.}
\end{figure}

The IRAC mosaicked images of GD\,61, CBS\,127, and HE\,1349$-$2305 all
reveal nearby background objects that have the potential to contaminate
aperture photometry (see Figure\,\ref{f-0435im} and Figure\,\ref{f-1349im}
for the images of GD\,61 and HE\,1349$-$2305 respectively). The background
object near GD\,61 is separated by $1\farcs9$ at P.A. $25\arcdeg$ in the
images taken with LIRIS (epoch 2011.2), and is readily resolved in the $H$
and $K$ bands. The near-infrared brightness of this neighbor is ($J,H,K$)
= (17.7,17.0,16.7)\,mag and never approaches the apparent magnitude of the
white dwarf. Based on this fact and the colors of the neighbor, it must be
a background object. This source contributes weakly in the 4.5\,$\mu$m
image, where it is well separated by \textsc{apex}, and is no longer seen
at 7.9\,$\mu$m. CBS\,127 and HE\,1349$-$2305 both have neighbors
$2\farcs8$ distant (north and west respectively); these are satisfactorily
separated by PSF fitting with \textsc{apex}.

\subsection{Spectral Energy Distributions}
\label{ss-sed}

Figures\,\ref{f-sed}--\ref{f-nond1} illustrate the spectral energy
distributions (SEDs) of the 15 metal-rich white dwarfs listed in
Table\,\ref{t-tar}. In addition to the supplementary near-infrared
photometry described above, we include shorter wavelength photometry from
a
variety of literature and catalog sources (e.g., \textit{GALEX};
\citealt{martinetal05-1}, SDSS DR7; \citealt{abazajianetal09-1}, DENIS;
\citealt{epchteinetal99-1}, and CMC; \citealt{cmc14}). In cases where
independent $JHK$ observations were not taken for this program, data from
2MASS \citep{skrutskieetal06-1} and UKIDSS \citep{lawrenceetal07-1} were
used. Where possible, a comparison of the IRAC photometry was made to data
available at similar wavelengths from the \textit{Wide-field Infrared
Survey Explorer} \citep[\textit{WISE}][]{wrightetal10-1}.

For flux scaling of the spectral models in the figures, the most reliable
photometry was used, which was primarily SDSS $ugriz$ and the $JHK$ data
taken specifically for this program. Specifically, the \textit{GALEX}
fluxes were not used to constrain the white dwarf atmospheric models
because 1) metal lines may significantly suppress these fluxes compared
with a pure helium (or hydrogen) model atmosphere and 2) the interstellar
reddening to these objects is not well known.

Table\,\ref{t-tar} lists the best available stellar parameters from the
literature for each of the science targets. We used these parameters as an
initial estimate for the white dwarf spectral models. A grid of synthetic
white dwarf atmospheric spectra from \citet{koester10-1} was interpolated
to these temperature and surface gravity estimates. After comparing the
fluxes from the models with the most reliable optical and near-infrared
data, in some cases we found that a slightly different effective
temperature provided a superior fit to the data (shown in the upper right
corner of the relevant figure). These parameters thus used in the SED fits
do not reflect an independent parameter determination, but rather our
attempt to best constrain the infrared photospheric emission. 

\subsection{Stars With An Infrared Excess}
\label{s-ire}

We model stars with an infrared excess as a combination of the white dwarf
photosphere and an optically thick, geometrically thin disk with a
temperature profile $T_{\rm disk} \propto r^{\beta}$, with $\beta=-3/4$
\citep{adamsetal88-1, chiang+goldreich97-1, jura03-1}. The ratio of
stellar radius to distance sets the absolute scale of the white dwarf
photospheric flux, and is fixed in our modeling based on the best
available parameters. The free parameters in the disk model are the inner
disk radius ($R_{\rm in}$), the outer disk radius ($R_{\rm out}$) and the
disk inclination ($i$). A grid of disk models was calculated with inner
($T_{\rm in}$) and outer disk temperatures ($T_{\rm out}$; each
corresponding to a value of $R_{\rm in}$ and $R_{\rm out}$) ranging from
100 to 1800\,K in steps of 50\,K, and inclination ranging from 0 to
90\arcdeg in steps of $5\arcdeg$. $T_{\rm out}$ was also fixed to be
cooler than $T_{\rm in}$. A least \chisq\ method was used to fit the $H$-,
$K$- and IRAC-band infrared fluxes and provide an estimate of the
uncertainties.

\begin{figure}
\begin{center}
\includegraphics[width=0.85\columnwidth]{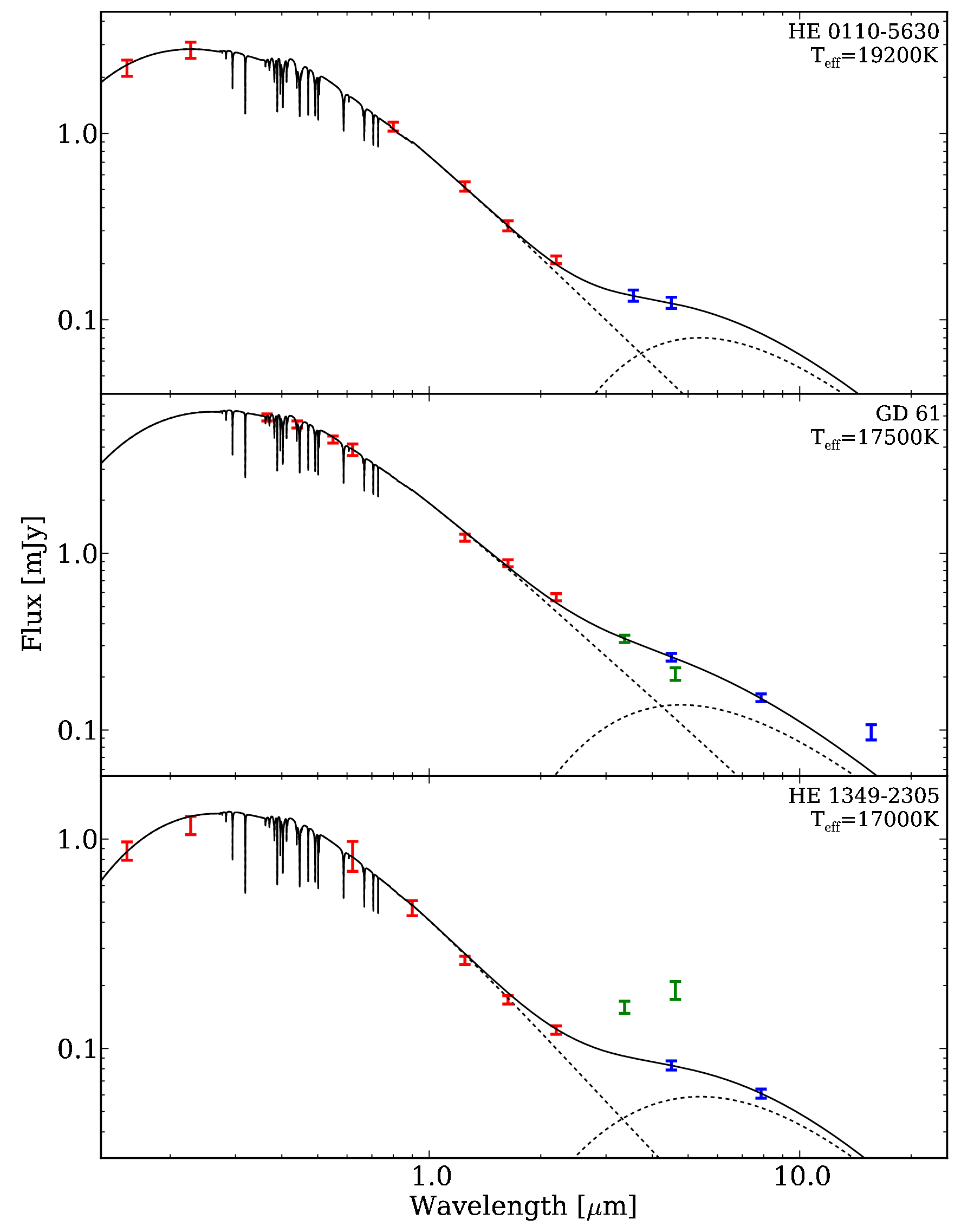}
\end{center}
\caption{\label{f-sed} SED of HE\,0110-5630, GD\,61 and HE\,1349$-$2305.
The short wavelength fluxes are: HE\,0110-5630: \textit{GALEX}, DENIS
$I$-band, and SOFI $JHK$, GD\,61: $UBV$ \citep{eggen68-1} CMC $r'$, and
LIRIS $JHK$, and HE\,1349$-$2305: \textit{GALEX}, CMC $r'$, DENIS $I$,
SOFI $JHK$. These are shown in red with error bars. \textit{WISE} data are
shown in green, however, they are clearly contaminated by the neighboring
source for HE\,1349$-$2305 (\S3.1). The IRAC and IRS fluxes are shown in
blue. The short wavelength photometry is fitted with a stellar model that
is displayed as a black dashed line. The minimum \chisq\ disk model fit,
shown in Figure\,\ref{f-chi2}, is displayed as a black dashed line, and
the sum of the two is shown in solid black.}
\end{figure}

\begin{figure}
\begin{center}
\vspace*{1cm}
\includegraphics[width=\columnwidth]{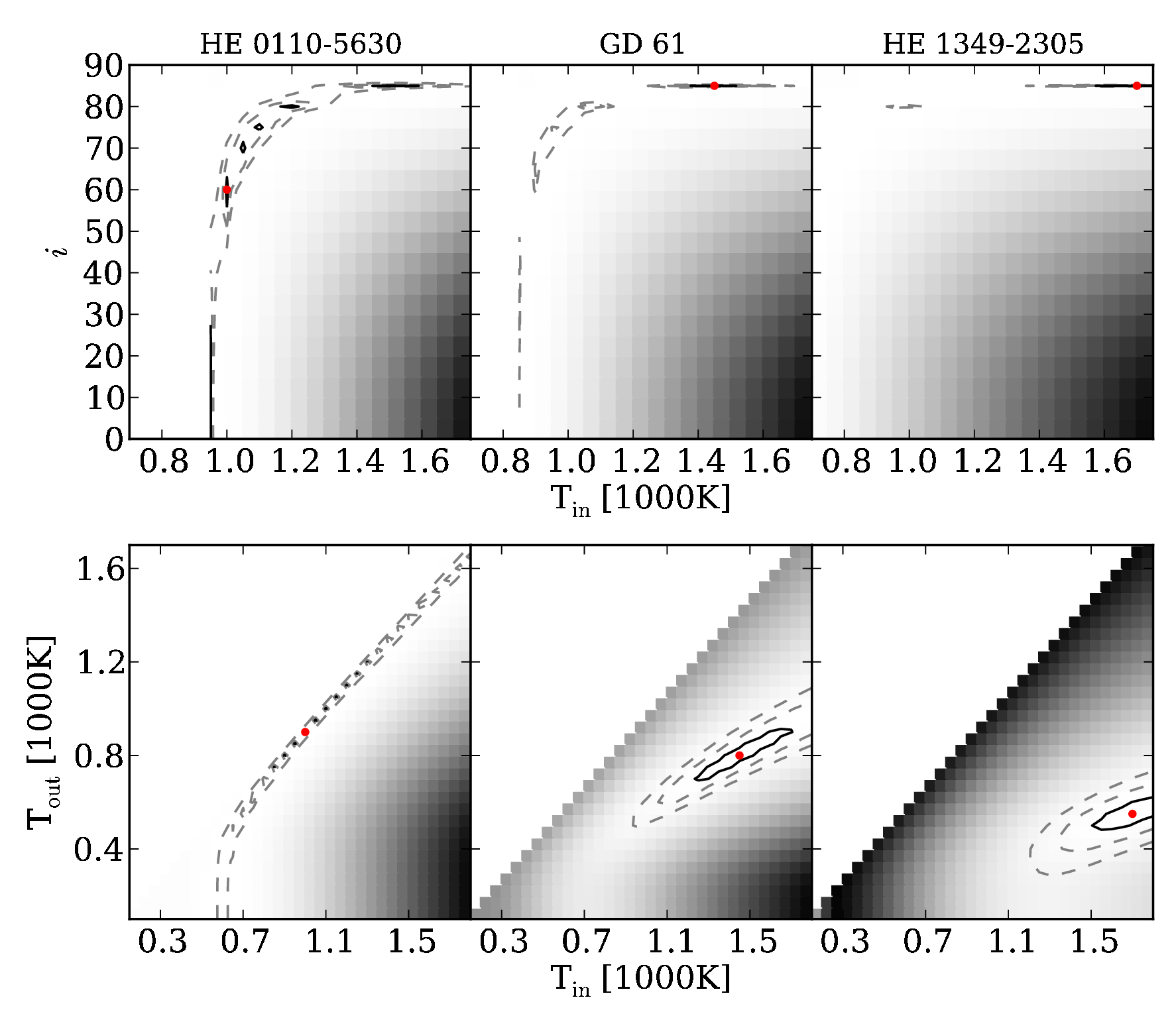}
\end{center}
\caption{\label{f-chi2} Disk modeling for HE\,0110-5630, GD\,61 and
HE\,1349$-$2305. The infrared excess was fitted as described in
\S\,\ref{s-ire}.
Briefly describing the format of the panels (see \S\,\ref{s-ire}); a slice
through the \chisq\ volume is displayed each of the pairs of upper and
lower panels for the minimum \chisq\ fit. The upper panel shows a slice at
the best fitting $T_{\rm out}$ and the lower panel is defined by the best
$i$. The best fitting solutions are: HE\,0110-5630: $T_{\rm in}=1,000$\,K,
$T_{\rm out}=900$\,K, $i=60\arcdeg$, GD\,61: $T_{\rm in}=1,450$\,K,
$T_{\rm out}=800$\,K, $i=85\arcdeg$ and HE\,1349$-$2305: $T_{\rm
in}=1,700$\,K, $T_{\rm out}=550$\,K, $i=85\arcdeg$. Regions of high
\chisq\ are shown as dark areas and the least \chisq\ solution is marked
as a red circle. Solid black lines show the $1\sigma$ contours around the
minimum, while the 2 and $3\sigma$ contours are shown as dashed gray
lines. Because only slices through the \chisq\ are shown, this does not
show the full extent of the \chisq\ surface. In the case of
HE\,0110-5630, $i$ vs $T_{\rm in}$, each of the regions enclosed by a
solid black line represents a parameter space where \chisq\ is within
$1\sigma$ of the minimum.
The \chisq\ method does not constrain the disk parameters very well for
HE\,0110-5630.}
\end{figure}

In Figure\,\ref{f-chi2}, the panels show a slice through the \chisq\ cube
at the best fit solution. The upper panel of each pair displays a slice at
the best fitting $T_{\rm out}$, whilst the best fit inclination defines
the slice in the lower panel. Regions of high \chisq\ are shown as dark
areas and the least \chisq\ solution is marked as a red circle and its
flux is displayed in Figure\,\ref{f-sed}. Solid black lines show the
$1\sigma$ contours around the minimum, while 2 and $3\sigma$ contours are
shown as dashed gray lines. The upper left of each lower panel is excluded
by the condition $T_{\rm in} > T_{\rm out}$.

The flat disk models naturally possess a modest degree of uncertainty.
The three free parameters ($T_{\rm in}$, $T_{\rm out}$ and $i$) are
somewhat degenerate in how they combine to determine the viewed solid
angle of the disk. Even in the case when longer wavelength data (such as
24\,$\mu$m photometry) are available, this degeneracy is not broken
\citep{juraetal07-1}. Broadly speaking, the inner disk edges are fairly
well constrained by their 2.2 and 3.6\,$\mu$m emission. The 4.5 to
7.9\,$\mu$m flux, however, can be reproduced by a relatively wide
temperature range and a higher inclination, or a narrower temperature
range and a lower inclination. Importantly, the newly detected disks mimic
the emission seen at more than one dozen dusty white dwarfs observed with
\textit{Spitzer} IRAC and MIPS \citep[e.g.][]{farihietal09-1} whose outer
disk radii are consistently within the Roche limit $d\gtrsim1$\,km solid
bodies ($\sim1\Rsun$). This corresponds to an outer disk temperature near
500\,K, which is a likely lower limit for $T_{\rm out}$ when interpreting
the \chisq\ surfaces. All together, the infrared emissions are precisely
that expected for dust particles resulting from the tidal destruction of a
large planetary body, whose constituent elements now rain onto and pollute
the surfaces of the host stars.

\textit{HE\,0110$-$5630}. The available photometry for HE\,0110$-$5630
consists of \textit{GALEX} far- and near-UV, DENIS $I$-band and $J$, $H$
and $K$ measurements taken with SOFI. The photospheric emission is
relatively well constrained by the $J$- and $H$-band fluxes and,
therefore, so is the extrapolation to longer infrared wavelengths. The
measured IRAC excess for HE\,0110$-$5630 is consistent with warm
circumstellar dust and is shown in Figure\,\ref{f-sed}. The best fit disk
model has an inner and outer temperature $1,000$ and $900$\,K,
respectively, at an inclination of $i=60\arcdeg$. This corresponds to a
narrow ring, however, as can be seen in the \chisq\ panels, the solution
is degenerate with respect to inner disk temperatures above 800\,K, and
inclination, and the outer temperature is not at all constrained. The
\chisq\ surfaces shown in Figure\,\ref{f-chi2}, show that a vast
range in disk temperatures could provide a fit to the excess. We therefore
cannot confidently estimate any of the disk parameters.

\textit{GD\,61}. From analysis of IRAC images and Peak-Up Imaging, this
object was found to have an excess in \citet{farihietal11-1}. Here we
include the $JHK$ data taken with LIRIS to better constrain the stellar
flux. These images resolve a nearby (background) source
(Figure\,\ref{f-0435im}) that almost certainly contaminates the 2MASS
photometry used in previous fits. From its spectral energy distribution
and brightness, this object is not associated with the white dwarf. In
the \textit{Spitzer} IRAC images, the PSF of the white dwarf and the
contaminant overlap slightly, however, the \textsc{apex} PSF-fitting
photometry cleanly separates the two objects. We can therefore confirm
the presence of a disk from the IRAC and IRS fluxes.

When performing \chisq\ fitting of the infrared excess, we exclude the
$15.6\,\mu$m flux due to a possible increase owing to silicate emission,
and therefore the disk continuum emission would be overestimated. Such
emission has been detected in all eight dusty white dwarfs observed
spectroscopically with IRS \citep{reachetal05-1, juraetal09-1}. A disk
model with inner and outer-disk temperatures of $1,450$ and $800$\,K,
respectively, and an inclination $i=85\arcdeg$ is the best fit to the
excess (Figure\,\ref{f-sed} and Figure\,\ref{f-chi2}). The disk model
temperature and inclination are again degenerate, and a large range of
parameters would fit the infrared fluxes. We therefore find the values
reported previously \citep[$T_{\rm in}=1,300$\,K, $T_{\rm out}=1,000$\,K
and $i=79\arcdeg$;][]{farihietal11-1} are not inconsistent with ours,
given the uncertainties in temperature and inclination.

\textit{HE\,1349$-$2305}. After deconvolving the white dwarf and
neighboring, background source (\S\,\ref{ss-bg}), a significant infrared
excess is still found over the stellar model (Figure\,\ref{f-sed}). The
white dwarf photosphere is well constrained by short wavelength and
near-infrared photometry shown in Table\,\ref{t-ind}. \textit{WISE} $3.4$
and $4.6\,\mu$m fluxes are also available for this object, however, the
nearby source seen in Figure\,\ref{f-1349im} is within the \textit{WISE}
beam width and therefore significantly contaminates the \textit{WISE}
fluxes. We therefore do not include these fluxes in the disk modeling. The
excess is best fitted with a disk having $T_{\rm in}=1,700$\,K, $T_{\rm
out}=550$\,K and an inclination of $i=85\arcdeg$. The uncertainties are
rather modest in this case.


\textit{NLTT\,51844}. This was one of two white dwarfs bright enough to
attempt blue Peak-Up image using IRS. The $15.6\,\mu$m flux measurement is
in excess over the white dwarf model by slightly more than $3\sigma$,
similar to the MIPS 24\,$\mu$m photometric excesses detected at G238-44
and G180-57 \citep{farihietal09-1}. \textit{Spitzer} galaxy counts at
these wavelengths are $6\times10^{-4}$ galaxies per square arc second
\citep{marleauetal04-1} for the brightness needed to produce an apparent
excess in these cases. Therefore the probability that such a source falls
within a radius of two IRS Blue Peak-Up FWHM ($3\farcs8$) can be as high
as 10\,per\,cent. Thus the probability that three sources have excesses
at roughly 20\,$\mu$m (and not at shorter wavelengths) due to background
objects, out of 30 stars surveyed at these longer wavelengths, can be as
high as 65\,per\,cent.

While we cannot yet rule out a chance alignment in these cases, the
astrometric position of NLTT\,51844 in the IRS image mosaic coincides
within $2\arcsec$ of its position in the IRAC images, based on the
absolute \textit{Spitzer} astrometry provided in the processed image
files. Both NLTT\,51844 and G180-57 are cool, helium-rich stars where
metals can be resident for Myr timescales, and hence their accretion
history is unconstrained. In any case, dust that emits only at
such long wavelengths must be sufficiently far from the white dwarf
that it cannot be the immediate source of the atmospheric metals. Still,
if the data for NLTT\,51844 (as well as G238-44 and G180-57) are not
spurious, we may be seeing an important clue to the nature of numerous
polluted white dwarfs without obvious infrared excesses.

\newpage

\subsection{Stars Without An Infrared Excess}
\label{s-nire}

\begin{figure}
\includegraphics[width=\columnwidth]{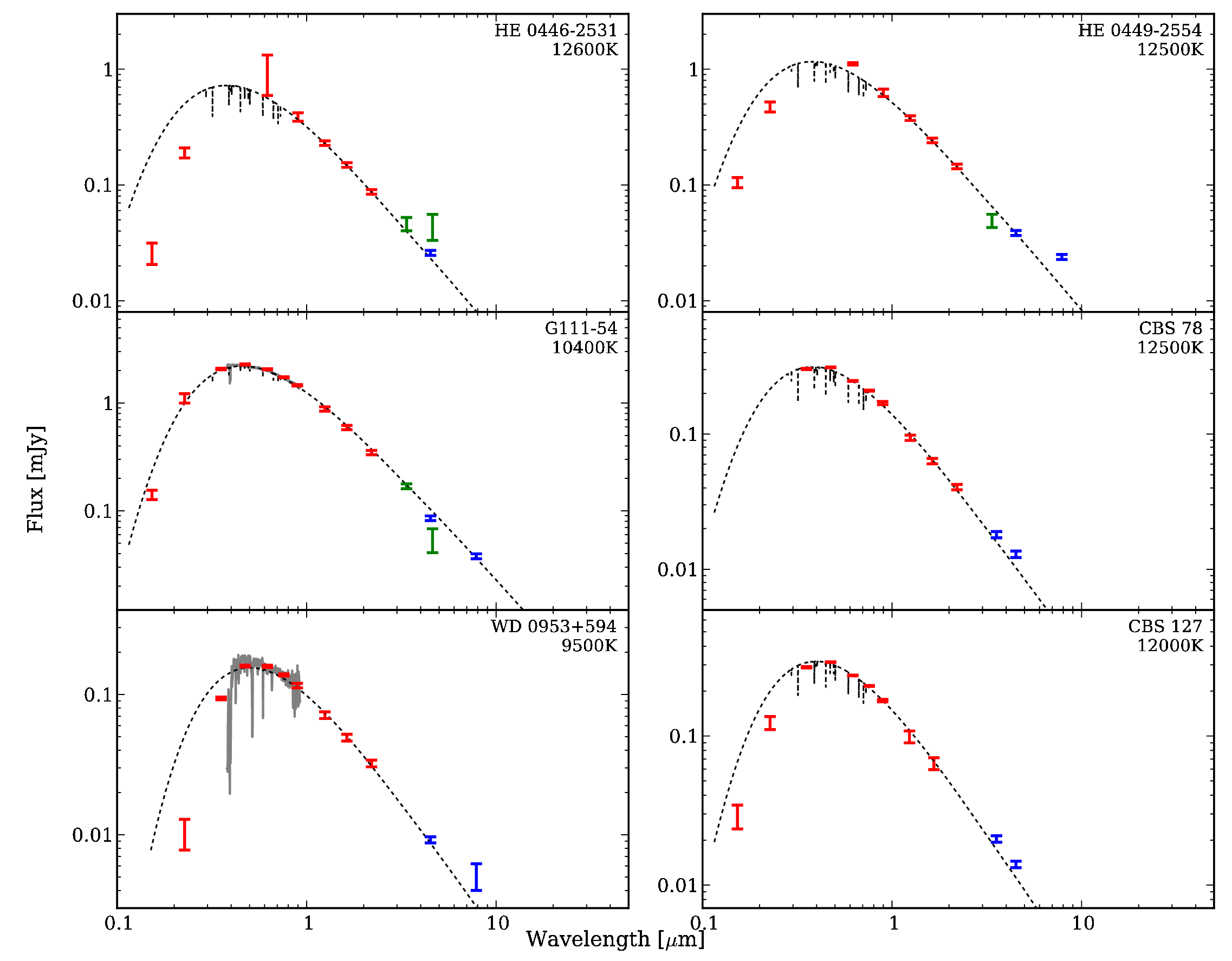}
\caption{\label{f-nond0} SEDs of 6 of the 12 science targets consistent
with photospheric emission (\S3.4). The figures follow the same general
format as Figure\,\ref{f-sed}. The SDSS optical spectrum of WD\,0953+954
is shown as a gray line. The infrared excess seen in the SED of
HE\,0449$-$2554 is caused by light from a background object.}
\end{figure}

\begin{figure}
\includegraphics[width=\columnwidth]{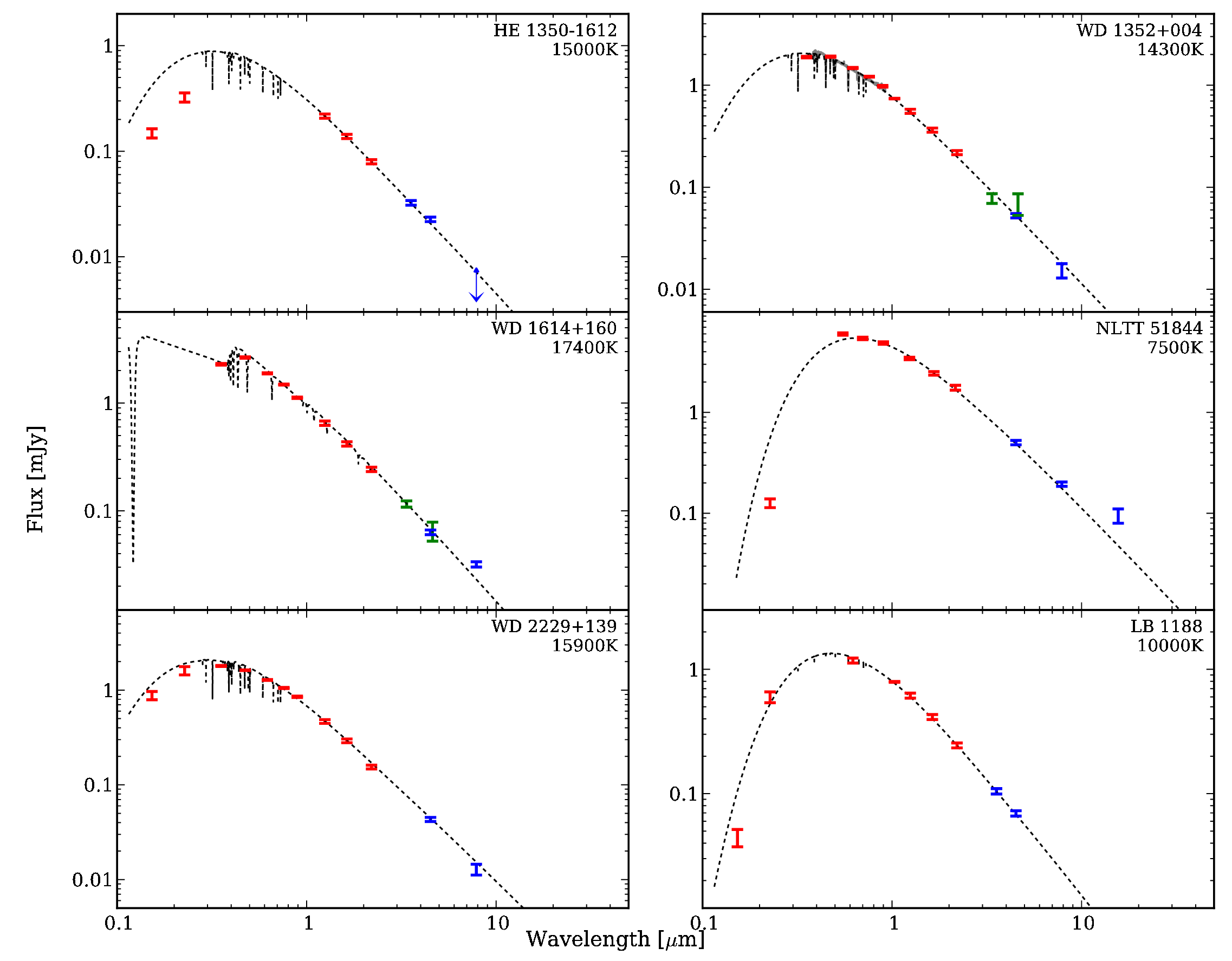}
\caption{\label{f-nond1} A continuation of Figure\,\ref{f-nond0} for
the remaining 6 of the 12 science targets consistent with photospheric
emission (\S3.4).}
\end{figure}

The remaining white dwarfs (Figures\,\ref{f-nond0} and \ref{f-nond1}) all
show no evidence for infrared excess emission. They are relatively well
constrained in the optical and near-infrared and therefore we can
extrapolate with some certainty to IRAC wavelengths. Below we provide
notes on a few particular stars.

\textit{HE\,0446$-$2531}. This star is perhaps the most highly polluted
white dwarf observed by \textit{Spitzer} and yet does not have an infrared
excess. If the analysis of \citet{friedrichetal00-1} is correct, this star
contains $8\times 10^{24}$\,g of metals in its atmosphere -- a mass
approaching that of Pluto. The long diffusion timescale for heavy
elements in this star allows for the possibility that a disk has been
fully dissipated, but the very large mass involved suggests we are still
within at most a few diffusion timescales of the accretion event. Two
possibilities are: 1) accretion is ongoing from gaseous debris, or 2)
accretion has ended but total mass involved was at least an order of
magnitude higher. It is noteworthy that this star and HS\,2253$+$8023 are
both highly polluted \citep{friedrichetal00-1} yet lack infrared excess,
with HE\,0446$-$2531 having nearly ten times more metal mass in its
convection zone. However, \citet{kleinetal11-1} found that HS\,2253$+$8023
has an order of magnitude less calcium than originally reported by
\citet{friedrichetal00-1}, and therefore the same may be true of
HE\,0446$-$2531.

\textit{HE\,0449$-$2554}. A partly-resolved extension is seen in the IRAC
7.9\,$\mu$m image, and the combined source is offset slightly from the
position of the white dwarf in the 4.5\,$\mu$m image. While the SOFI $JHK$
images of this star do not reveal any sources within a few arc seconds of
the science target, the marginal, apparent excess seen in the SED
(Figure\,\ref{f-nond0}) is probably caused by a background object. Better
data are needed to rule out contamination in IRAC.

\textit{CBS\,127}. As mentioned above, a background object lies within a
few arc seconds of the white dwarf, and could potentially contaminate
moderate-size apertures used for IRAC photometry. We used PSF-fitting
routines within \textsc{apex} to ensure both stars were photometrically
disentangled. There is no infrared excess at this white dwarf.

\textit{HE\,1350$-$1612} was observed in both the Cycle\,5 and 6 programs
and images were thus taken at 3.6, 4.5 (twice), and 7.9\,$\mu$m. The
average 4.5\,$\mu$m flux is shown in Table\,\ref{t-tar} and
Figure\,\ref{f-nond1}. The white dwarf was not detected in the 7.9\,$\mu$m
image, and thus a $3\sigma$ upper limit is given. Nevertheless, all the
infrared data are consistent with the predicted white dwarf photosphere.

\textit{PHL\,131} is a 40,000\,K white dwarf with interstellar calcium
absorption, and hence not metal-rich. We do not analyze this star but its
IRAC fluxes are reported in Table\,\ref{t-flu}.

\section{Discussion}
\label{s-dis}

\subsection{Updated Statistics and Accretion Rates}
\label{ss-dbz}

Of the 15 metal-polluted stars surveyed, three have an infrared excess
consistent with circumstellar dust disk, and indicative of a remnant
planetary system. Including the objects presented here, the total number
of metal-polluted white dwarfs observed with \textit{Spitzer} in Cycle\,1
through 7 is 92. This includes the 52 objects from Table\,4 of
\citet{farihietal09-1}, 8 from Table\,1 of \citet{farihietal10-1}, 3 white
dwarfs with gas disks from \citet{gaensickeetal06-3, gaensickeetal07-1,
gaensickeetal08-1}, 14 DBZ white dwarfs from \citet{xu+jura12-1} and the
15 objects studied here. Of these 92 objects, DAZ and DBZ-type white
dwarfs are represented in proportions of 39 and 53 respectively.

Our expanded study corroborates the previous finding
\citep{kilicetal08-1, farihietal09-1, farihietal10-1} that the detection
of infrared excess is less frequent among DBZ white dwarfs than among
their DAZ counterparts. 11 of 38 (29\,per\,cent) surveyed DAZ stars have
circumstellar dust, whereas the same fraction for surveyed DBZ-type white
dwarfs is only 7 of 52 (13\,per\,cent)\footnote{We exclude here the two
DAZ and one DBZ that were already known to have gaseous disks
\citep{gaensickeetal08-1} prior to their \textit{Spitzer} observations.
The stars discussed in \citet{farihietal11-3} were also not included
because they were observed with \textit{Spitzer} after they were found to
have an infrared excess in \citet{girvenetal11-1}.}. Because one would not
expect disks to preferentially form around stars based on their
atmospheric properties, the difference must be due to the longer diffusion
timescales \citep{paquetteetal86-1, koester+wilken06-1}. We will return to
 this point in more detail below.

\begin{figure}
\begin{center}
\includegraphics[width=\columnwidth]{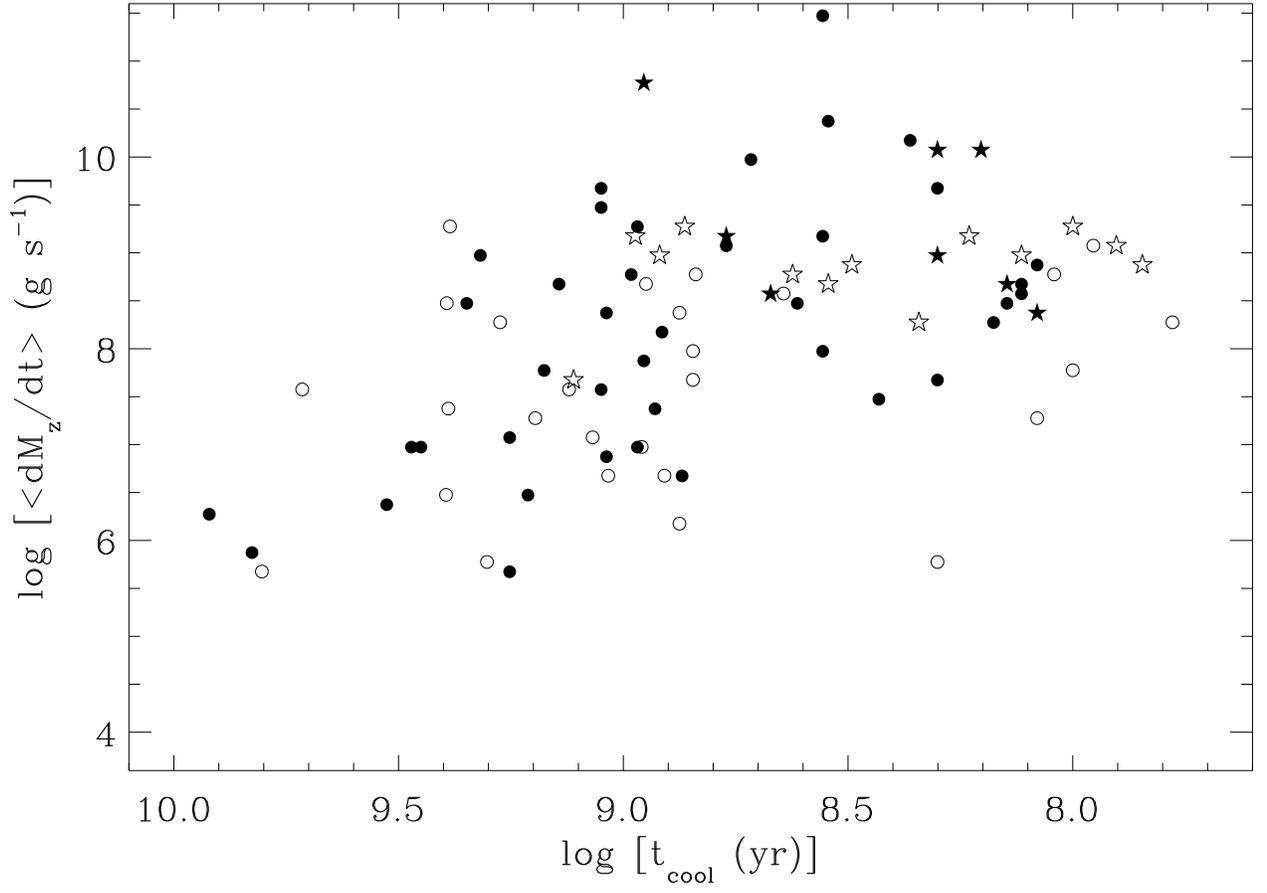}
\end{center}
\caption{\label{f-dMdt} Time-averaged dust accretion rates vs. cooling age
of the 88 metal-polluted white dwarfs, with published abundances, observed
with \textit{Spitzer} IRAC in Cycle 1 through 7. DAZ and DBZ-type stars
are plotted as open and filled circles respectively, while objects with
infrared excess are displayed as stars rather than circles.}
\end{figure}

Figure\,\ref{f-dMdt} illustrates the time-averaged metal accretion rate
versus cooling age for the 88 metal-rich white dwarfs now observed by
\textit{Spitzer} with published abundances (four of \citealt{xu+jura12-1}
stars, including one with a detected infrared excess, do not have
published abundances), updating Figure\, 10 from \citet{farihietal10-1}.
All accretion rates were calculated based on the accumulated metal
abundances, using Equation\,2 of \citet{koester+wilken06-1}, as performed
in \citet{farihietal09-1}.
One difference between the method used in \citet{farihietal09-1} and here
is that in previous studies instantaneous and time-averaged accretion
rates were calculated assuming the in-falling material either had Solar
calcium abundances \citep[$1:43$;][]{koester+wilken06-1}, or
$1$\,per\,cent Solar \citep[$1:109$; i.e. metals only;][]{juraetal07-1,
farihietal09-1} by mass fraction, based on the measured calcium abundance.

While the latter approach is still likely to be broadly correct, recent
progress in the field enables a more accurate estimation. There are now of
order 10 metal-polluted white dwarfs with measured Mg, Si, Ca, and Fe
abundances \citep{zuckermanetal03-1, zuckermanetal07-1, zuckermanetal10-1,
zuckermanetal11-1}, and a handful also with O abundances
\citep{dufouretal10-1, kleinetal10-1, vennesetal10-1, kleinetal11-1,
melisetal11-1, vennesetal11-1}. Based on these data, it is increasingly
clear that the accreted material has a composition similar to that of
rocky, terrestrial material of the Solar System. \citet{zuckermanetal10-1}
find that Ca represents, on average, close to 1 part in 60, by mass, of
all the accreted heavy elements. For comparison, this ratio is 1 part in
62.5 for the Earth \citep{allegreetal95-1}.
Based on these facts, Figure\,\ref{f-dMdt} and Table\,\ref{t-tar} employ
accretion rates, and convective envelope masses, assuming Ca is
$1.6$\,per\,cent ($1/62.5$) of the total mass. To convert between the
accretion rates quoted in \citet{farihietal09-1, farihietal10-1} and those
shown in Figure\,\ref{f-dMdt}, one must multiply by a factor of $109/62.5$
for DAZ white dwarfs and $43/62.5$ for DBZ white dwarfs.

Compared to the same Figures shown in previous (related) papers, the
accretion rates for the DAZ white dwarfs are therefore slightly decreased,
while their DBZ counterparts are slightly increased.

\subsection{A Simple Estimate of the Disk Lifetime}
\label{ss-lt}

The mass of metals in the convective envelope of DBZ-type white dwarfs
provides a lower limit on the total mass of the destroyed parent body.
From the DAZ stars, we can infer the instantaneous accretion rates because
one can safely assume accretion-diffusion equilibrium. There is no
a priori reason that disks around hydrogen and helium-rich stars
should be different \citep{juraetal07-1}. Therefore, assuming the inferred
accretion rates of DAZ stars represent an accurate cross section over disk
lifetimes, and similarly for the observed metal masses contained within
DBZ stellar envelopes, we can obtain a typical disk lifetime by combining
the average of these two observed quantities, i.e. 

\begin{equation}
t_{\rm disk}\sim \frac{\langle M_z \rangle_{\rm DBZ,disk}}{\langle dM_z/dt
\rangle_{\rm DAZ,disk}}
\end{equation}

\begin{figure}
\begin{center}
\includegraphics[width=\columnwidth]{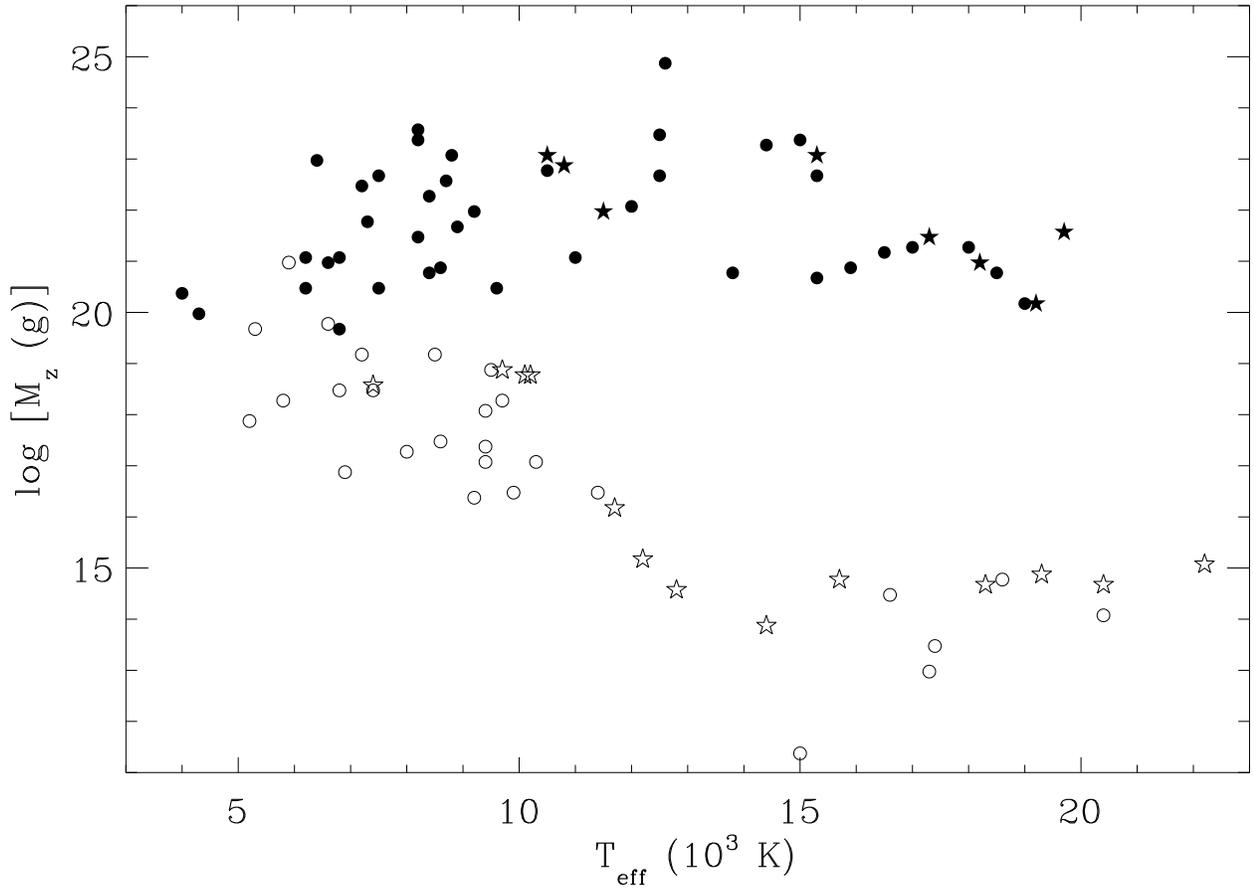}
\end{center}
\caption{\label{f-Mzt} Mass of metals within the convective envelopes (or
above $\tau=5$, whichever is larger) of the 88 metal-polluted white
dwarfs, with published abundances, observed with \textit{Spitzer} IRAC in
Cycle 1 through 7, plotted as a function of effective stellar temperature.
DAZ and DBZ-type stars are plotted as open and filled circles
respectively, while objects with infrared excess are displayed as stars
rather than circles.}
\end{figure}

Figure\,\ref{f-Mzt} demonstrates that the mass of metals in the convective
envelope of DBZ white dwarfs is not a strong function of temperature, and
hence cooling age. Therefore, calculating the average convective envelope
metal mass across cooling ages should be robust. In contrast, the
convective envelope depth increases substantially in this temperature
range for DAZ stars \citep{koester09-1}, as can be seen by the large
increase in convective envelope metal masses at longer cooling ages in
Figure\,\ref{f-Mzt}.

Among 13 DAZ white dwarfs observed to have infrared excess with
\textit{Spitzer}, the average metal accretion rate is
$9.7\times10^8$\,g\,s$^{-1}$. Among 8 DBZ-type stars with dust detected by
\textit{Spitzer}\footnote{This does not include
SDSS\,J220934.84$+$122336.5, the DBZ star from \citet{xu+jura12-1} with an
infrared excess, because no calcium abundances are published.} the
average metal content of the convection zone is $4.1\times 10^{22}$\,g,
and we thus estimate $t_\mathrm{disk}\simeq1.3\times10^6$\,yr. However,
Figure\,\ref{f-dMdt} shows that the accretion rates and convection zone
metal content vary by orders of magnitude, and suggests that using the
logarithmic mean and standard deviation, $\langle
\log[dM_z/dt$\,(g\,s$^{-1})]\rangle=8.8\pm0.4$ and $\langle
\log[M_z$\,(g)]$\rangle =21.9\pm1.1$, may be a more appropriate
choice. The resulting estimate for the disk lifetime is $\log[t_{\rm
disk}$\,(yr)$]=5.6\pm1.1$. Because the metal masses in the DBZ stars are
lower limits on the total parent body masses, it is reasonable to presume
the same for the disk lifetimes.

At least 1\,per\,cent of all white dwarfs with cooling ages less than
0.5\,Gyr have dusty disks \citep{farihietal09-1, girvenetal11-1,
steeleetal11-1}. Assuming that all white dwarfs host (remnants of)
planetary systems and go through intermittent phases of debris, one would
expect any given star to exhibit detectable infrared excess for
$0.01\times0.5\times10^9$\,yr $=5\times10^6$\,yr, which is broadly
consistent with our estimate above.

\subsection{How Complex is the Evolution of the Dust Disks?}

Different possible scenarios for the evolution of dust disks, and
corresponding estimates for their lifetimes have been discussed in more
detail by \citet{jura08-1}, \citet{rafikov11-1,rafikov11-2}, and
\citet{bochkarev+rafikov01-1}. 

A main motivation of the work of \citet{jura08-1} was to explain the
existence of metal-polluted white dwarfs without infrared excess detection
by the continuous accretion from a gaseous disk that is replenished by the
repeated tidal destruction of multiple small asteroids. In this scenario,
dust disks are associated with the disruption of massive asteroids, and
\citet{jura08-1} estimated the lifetimes of the gaseous and dust disks to
be $\sim0.5\times10^4$\,yr and $\sim1.5\times10^5$\,yr, respectively,
though he underlined the uncertainty of these estimates. \citet{jura08-1}
discussed his scenario in the context of a much smaller sample of
metal-polluted white dwarfs, in which the majority of high $dM/dt$ stars
were displaying infrared excess identifying the presence of dusty disks.
It is unlikely that the much larger accretion rates that are now known for
some stars without infrared excess can be explained by multiple impacts
of small asteroids.  

\citet{rafikov11-1} showed that Poynting-Robertson drag can explain
accretion rates of $\sim10^8$\,g/s, and estimated the lifetime of massive
($\sim10^{22}$\,g) dust disks to several Myr. The higher accretion rates
observed in a number of white dwarfs require the presence of gas that
increases the viscosity, and \citet{rafikov11-2} shows that sublimation of
the inner disk may be sufficient to lead to a runaway evolution which
produces accretion rates of $10^{10}-10^{11}$\,g/s, exhausting the disk in
$\sim10^5$\,yr. \citet{rafikov11-2} explicitly excluded the generation of
gas by multiple impacts of additional asteroids. Extending the analysis to
the global evolution of the debris disks, \citet{bochkarev+rafikov01-1}
confirm that massive disks have expected life times of several Myr if
subject to Poynting-Robertson drag only, but that sublimation may
substantially shorten these estimates. 

Our disk lifetime estimate in the previous section is broadly compatible
with all the estimates summarized above. However, our approach assumed
that accretion rates do not change significantly as a function of disk
age, and that the \textit{Spitzer} sample is representative of the
potential changes in the accretion rate. \citet{zuckermanetal10-1}
suggested that accretion onto a white dwarf could proceed through three
distinct phases; a build-up, a steady state, and a decline, where the
steady state lifetime should be orders of magnitude longer than the
beginning and ending episodes. If correct, and disk lifetimes are
typically $\simeq10^6$\,yr, then it will be very unlikely for an arbitrary
white dwarf with circumstellar dust to be observed in anything other than
a steady state phase.

Inspection of Figure\,\ref{f-dMdt} shows, as already mentioned in
\S\,\ref{ss-dbz}, that the fraction of DBZ with \textit{Spitzer} infrared
excess is substantially smaller than that of DAZ. The detection of dust
seems to be associated with a minimum accretion rate, and if we only
consider white dwarfs with $dM/dt\ge10^8\mathrm{g\,s^{-1}}$, we find that
10/21 (48\,per\,cent) of DAZ have infrared excess, compared to only 7/30
(23\,per\,cent) for the DBZ\footnote{Again excluding the two DAZ and one
DBZ that were identified because of their gaseous discs
\citep{gaensickeetal08-1}}. This difference is, however, subject to small
number statistics and future observations may show that both DAZ and DBZ
are equally likely to show an infrared excess. If there is a real
disparity, this fact is most likely related to the one significant
difference between DBZ and DAZ: their diffusion time scales, and allows
some independent gauge of the disk lifetime.
Assuming a typical lifetime of the disks that is similar to or greater
than the diffusion time scale of the DBZ stars ($t_{\rm disk} \gtrsim
t_{\rm diff}\simeq10^5-10^6$), we would expect a similar ratio of DAZ and
DBZ stars exhibiting infrared excess. By contrast, if $t_{\rm disk} \ll
t_{\rm diff}$, it would be very unlikely to detect the disks around any
given DBZ. Thus, the small but non-negligible fraction of DBZ with
observed infrared excess leads us to conclude that the typical disk life
time is somewhat shorter than the diffusion time scales in DBZ~--~or that
the disk lifetimes vary by substantial amounts (which is in fact
consistent with the large uncertainty in our logarithmic estimate of the
life time). It appears unlikely that a large fraction of the highly
polluted DBZ white dwarfs without infrared excess are currently accreting
from a gaseous disk, as the life times of these gaseous disks are short
\citep{jura08-1}.

Figure\,\ref{f-dMdt} also reveals a fairly well defined upper limit in
$dM/dt$ for DAZ white dwarfs. Based on our updated calculation of the
accretion rates, there are no DAZ white dwarfs observed with
\textit{Spitzer} with accretion rates above $2 \times 10^{9}$\,gs$^{-1}$.
The most extremely metal-polluted DAZ white dwarf star currently known,
GALEX\,1931$+$0117\footnote{Not included in our Figure because it has not
been observed with \textit{Spitzer}} \citep{vennesetal10-1}, still only
accretes at a rate of $3-4 \times 10^{9}$\,gs$^{-1}$
\citep{melisetal11-1, vennesetal11-1}. In contrast, there are several DBZ
white dwarfs with metal accretion rates one to two orders of magnitude
larger than observed in any DAZ, and substantially exceeding the rates
that can be explained by Poynting-Robertson drag \citep{rafikov11-1,
rafikov11-2}. For instance, HE\,0446--2531, HE\,0449--2554, and
HE\,1350--162 all have accreted several $10^{23}$\,g to nearly
$10^{25}$\,g within the last diffusion time scale, i.e. $\simeq10^6$\,yr
(or correspondingly larger amounts of material if the accretion ended
several diffusion time scales ago). This is comparable to the extremely
metal-polluted DZ white dwarf, SDSS\,J0956$+$5912, which has at least
$1.5 \times 10^{23}$\,g of metals in its
atmosphere\citep{koesteretal11-1}.
Assuming the planetary bodies and physical mechanisms do not depend on the
type of white dwarf, this substantial difference in the maximum accretion
rates found for DAZ and DBZ white dwarfs forces us to conclude that either
the calculated metal accretion rates for DBZ white dwarfs suffer from
systematic errors, or the extremely high metal accretion rates found among
DBZ white dwarfs simply have not been seen yet in DAZ white dwarfs.

One may speculate that both DAZ and DBZ white dwarfs can undergo a
short-lived phase of very high rate accretion. Whereas the long diffusion
time scales of DBZ provide an efficient ``memory'' to such events, the
opposite is true for the DAZ, where the large amount of metals would be
cleared out of the atmosphere within days to at most years~--~reducing the
probability of witnessing such an event. A number of possible scenarios
that would lead to such phases are conceivable. Secondary impacts of
asteroids on an existing massive dust disk could generate sufficient
amounts of gas to lead to runaway accretion \citep{jura08-1}.
Alternatively, a collisional cascade in the initially highly eccentric
disk may generate sufficient amounts of gas to cause runaway accretion,
effectively preventing the formation of a longer lived disk. In either
scenario, the perturbed debris disks are expected to evolve on relatively
short timescales, and \citet{gaensickeetal08-1} have shown that the
structure of the gaseous disk in SDSS\,J084539.17+225728.0 changes on
timescales of a few years. An initial spike in the accretion rate could
also be related to the direct accretion of a substantial fraction of the
disrupted asteroid, with the possible subsequent formation of a debris
disk feeding the white dwarf at a lower rate. Another alternative comes
from the fact that asteroid accretion events are, by definition, produced
from eccentric orbits. One can therefore envisage that a fraction of disk
are therefore born in an unstable configuration. The disk would, similarly
to the above scenario, have many disk--disk interactions, producing a
significant amount of gas. Therefore the disk could accreted at a very
high rate, leaving a heavily-polluted white dwarf without a dust disk.

\section{Conclusions}
\label{s-con}

We obtained and analyzed \textit{Spitzer} observations of 15 white dwarfs
with metal-polluted atmospheres, all but one having helium-dominated
atmospheres. Of these, HE\,0110$-$5630, GD\,61 and HE\,1349$-$2305 are
found to have an infrared excess consistent with closely-orbiting
circumstellar dust. These disks are likely formed from the disruption of
large asteroid analogs within the remnant planetary systems. A marginal
excess is measured at $15.6\, \mu$m around NLTT\,51844, but more data are
needed to rule out contamination from extragalactic sources.

This survey nearly doubles the number of disk detections around DBZ white
dwarfs. Using this substantially enlarged sample, we estimate a typical
disk lifetime by comparing the accreted metal masses inferred from DBZ
stars with dust to the instantaneous accretion rates for DAZ stars with
dust. Accounting for the large scatter by taking the logarithmic average,
we find $\log[t_\mathrm{disk}\mathrm{(yr)}]=5.6\pm1.1$, which is
compatible with the relatively large range of disk lifetimes estimated
from different theoretical models of white dwarf disks.

The fraction of highly-polluted DBZ white dwarfs exhibiting an infrared
excess is low (23\,per\,cent) compared to that among DAZ white dwarfs
(48\,per\,cent). Assuming that the formation and evolution of
circumstellar disks is similar for both types of stars, this difference
suggests that the disk life times are typically shorter than the DBZ
diffusion time scales.

We also show that the highest time-averaged accretion rates are found
among white dwarfs with helium-rich atmospheres, many of which do not
exhibit infrared excess, and we suggest that these stars have experienced
very high accretion rates during short-lived phases. These events should
occur in hydrogen-dominated white dwarfs as well, but their short
diffusion timescales substantially lower the probability of detection. 

\section*{Acknowledgements}
\label{s-ack}

This work is based on observations made with the \textit{Spitzer Space
Telescope}, which is operated by the Jet Propulsion Laboratory, Caltech,
under NASA contracts 1407 and 960785. Some data presented herein are part
of the Sloan Digital Sky Survey, which is managed by the Astrophysical
Research Consortium for the Participating Institutions
(http://www.sdss.org/). This work makes use of data products from the Two
Micron All Sky Survey, which is a joint project of the University of
Massachusetts and IPAC/Caltech, funded by NASA and the NSF. A small part
of the work uses data products from the \textit{Wide-field Infrared Survey
Explorer}, which is a joint project of the UCLA, and JPL/Caltech, funded
by the NASA.



\begin{thebibliography}{77}
\expandafter\ifx\csname natexlab\endcsname\relax\def\natexlab#1{#1}\fi

\bibitem[{{Abazajian} {et~al.}(2009){Abazajian}, {Adelman-McCarthy},
  {Ag{\"u}eros}, {Allam}, {Allende Prieto}, {An}, {Anderson}, {Anderson},
  {Annis}, {Bahcall}, {Bailer-Jones}, {Barentine}, {Bassett}, {Becker},
  {Beers}, {Bell}, {Belokurov}, {Berlind}, {Berman}, {Bernardi},
{Bickerton},
  {Bizyaev}, {Blakeslee}, {Blanton}, {Bochanski}, {Boroski},
{Brewington},
  {Brinchmann}, {Brinkmann}, {Brunner}, {Budav{\'a}ri}, {Carey},
{Carliles},
  {Carr}, {Castander}, {Cinabro}, {Connolly}, {Csabai}, {Cunha},
{Czarapata},
  {Davenport}, {de Haas}, {Dilday}, {Doi}, {Eisenstein}, {Evans},
{Evans},
  {Fan}, {Friedman}, {Frieman}, {Fukugita}, {G{\"a}nsicke}, {Gates},
  {Gillespie}, {Gilmore}, {Gonzalez}, {Gonzalez}, {Grebel}, {Gunn},
  {Gy{\"o}ry}, {Hall}, {Harding}, {Harris}, {Harvanek}, {Hawley},
{Hayes},
  {Heckman}, {Hendry}, {Hennessy}, {Hindsley}, {Hoblitt}, {Hogan},
{Hogg},
  {Holtzman}, {Hyde}, {Ichikawa}, {Ichikawa}, {Im}, {Ivezi{\'c}},
{Jester},
  {Jiang}, {Johnson}, {Jorgensen}, {Juri{\'c}}, {Kent}, {Kessler},
{Kleinman},
  {Knapp}, {Konishi}, {Kron}, {Krzesinski}, {Kuropatkin}, {Lampeitl},
  {Lebedeva}, {Lee}, {Lee}, {Leger}, {L{\'e}pine}, {Li}, {Lima}, {Lin},
{Long},
  {Loomis}, {Loveday}, {Lupton}, {Magnier}, {Malanushenko},
{Malanushenko},
  {Mandelbaum}, {Margon}, {Marriner}, {Mart{\'{\i}}nez-Delgado},
{Matsubara},
  {McGehee}, {McKay}, {Meiksin}, {Morrison}, {Mullally}, {Munn},
{Murphy},
  {Nash}, {Nebot}, {Neilsen}, {Newberg}, {Newman}, {Nichol}, {Nicinski},
  {Nieto-Santisteban}, {Nitta}, {Okamura}, {Oravetz}, {Ostriker}, {Owen},
  {Padmanabhan}, {Pan}, {Park}, {Pauls}, {Peoples}, {Percival}, {Pier},
{Pope},
  {Pourbaix}, {Price}, {Purger}, {Quinn}, {Raddick}, {Fiorentin},
{Richards},
  {Richmond}, {Riess}, {Rix}, {Rockosi}, {Sako}, {Schlegel}, {Schneider},
  {Scholz}, {Schreiber}, {Schwope}, {Seljak}, {Sesar}, {Sheldon},
{Shimasaku},
  {Sibley}, {Simmons}, {Sivarani}, {Smith}, {Smith}, {Smol{\v c}i{\'c}},
  {Snedden}, {Stebbins}, {Steinmetz}, {Stoughton}, {Strauss}, {Subba
Rao},
  {Suto}, {Szalay}, {Szapudi}, {Szkody}, {Tanaka}, {Tegmark}, {Teodoro},
  {Thakar}, {Tremonti}, {Tucker}, {Uomoto}, {Vanden Berk}, {Vandenberg},
  {Vidrih}, {Vogeley}, {Voges}, {Vogt}, {Wadadekar}, {Watters},
{Weinberg},
  {West}, {White}, {Wilhite}, {Wonders}, {Yanny}, {Yocum}, {York},
{Zehavi},
  {Zibetti}, \& {Zucker}}]{abazajianetal09-1}
{Abazajian}, K.~N., {Adelman-McCarthy}, J.~K., {Ag{\"u}eros}, M.~A.,
{Allam},
  S.~S., {Allende Prieto}, C., {An}, D., {Anderson}, K.~S.~J.,
{Anderson},
  S.~F., {Annis}, J., {Bahcall}, N.~A., {Bailer-Jones}, C.~A.~L.,
{Barentine},
  J.~C., {Bassett}, B.~A., {Becker}, A.~C., {Beers}, T.~C., {Bell},
E.~F.,
  {Belokurov}, V., {Berlind}, A.~A., {Berman}, E.~F., {Bernardi}, M.,
  {Bickerton}, S.~J., {Bizyaev}, D., {Blakeslee}, J.~P., {Blanton},
M.~R.,
  {Bochanski}, J.~J., {Boroski}, W.~N., {Brewington}, H.~J.,
{Brinchmann}, J.,
  {Brinkmann}, J., {Brunner}, R.~J., {Budav{\'a}ri}, T., {Carey}, L.~N.,
  {Carliles}, S., {Carr}, M.~A., {Castander}, F.~J., {Cinabro}, D.,
{Connolly},
  A.~J., {Csabai}, I., {Cunha}, C.~E., {Czarapata}, P.~C., {Davenport},
  J.~R.~A., {de Haas}, E., {Dilday}, B., {Doi}, M., {Eisenstein}, D.~J.,
  {Evans}, M.~L., {Evans}, N.~W., {Fan}, X., {Friedman}, S.~D.,
{Frieman},
  J.~A., {Fukugita}, M., {G{\"a}nsicke}, B.~T., {Gates}, E., {Gillespie},
B.,
  {Gilmore}, G., {Gonzalez}, B., {Gonzalez}, C.~F., {Grebel}, E.~K.,
{Gunn},
  J.~E., {Gy{\"o}ry}, Z., {Hall}, P.~B., {Harding}, P., {Harris}, F.~H.,
  {Harvanek}, M., {Hawley}, S.~L., {Hayes}, J.~J.~E., {Heckman}, T.~M.,
  {Hendry}, J.~S., {Hennessy}, G.~S., {Hindsley}, R.~B., {Hoblitt}, J.,
  {Hogan}, C.~J., {Hogg}, D.~W., {Holtzman}, J.~A., {Hyde}, J.~B.,
{Ichikawa},
  S.-i., {Ichikawa}, T., {Im}, M., {Ivezi{\'c}}, {\v Z}., {Jester}, S.,
  {Jiang}, L., {Johnson}, J.~A., {Jorgensen}, A.~M., {Juri{\'c}}, M.,
{Kent},
  S.~M., {Kessler}, R., {Kleinman}, S.~J., {Knapp}, G.~R., {Konishi}, K.,
  {Kron}, R.~G., {Krzesinski}, J., {Kuropatkin}, N., {Lampeitl}, H.,
  {Lebedeva}, S., {Lee}, M.~G., {Lee}, Y.~S., {Leger}, R.~F.,
{L{\'e}pine}, S.,
  {Li}, N., {Lima}, M., {Lin}, H., {Long}, D.~C., {Loomis}, C.~P.,
{Loveday},
  J., {Lupton}, R.~H., {Magnier}, E., {Malanushenko}, O., {Malanushenko},
V.,
  {Mandelbaum}, R., {Margon}, B., {Marriner}, J.~P.,
{Mart{\'{\i}}nez-Delgado},
  D., {Matsubara}, T., {McGehee}, P.~M., {McKay}, T.~A., {Meiksin}, A.,
  {Morrison}, H.~L., {Mullally}, F., {Munn}, J.~A., {Murphy}, T., {Nash},
T.,
  {Nebot}, A., {Neilsen}, E.~H., {Newberg}, H.~J., {Newman}, P.~R.,
{Nichol},
  R.~C., {Nicinski}, T., {Nieto-Santisteban}, M., {Nitta}, A., {Okamura},
S.,
  {Oravetz}, D.~J., {Ostriker}, J.~P., {Owen}, R., {Padmanabhan}, N.,
{Pan},
  K., {Park}, C., {Pauls}, G., {Peoples}, J., {Percival}, W.~J., {Pier},
J.~R.,
  {Pope}, A.~C., {Pourbaix}, D., {Price}, P.~A., {Purger}, N., {Quinn},
T.,
  {Raddick}, M.~J., {Fiorentin}, P.~R., {Richards}, G.~T., {Richmond},
M.~W.,
  {Riess}, A.~G., {Rix}, H.-W., {Rockosi}, C.~M., {Sako}, M., {Schlegel},
  D.~J., {Schneider}, D.~P., {Scholz}, R.-D., {Schreiber}, M.~R.,
{Schwope},
  A.~D., {Seljak}, U., {Sesar}, B., {Sheldon}, E., {Shimasaku}, K.,
{Sibley},
  V.~C., {Simmons}, A.~E., {Sivarani}, T., {Smith}, J.~A., {Smith},
M.~C.,
  {Smol{\v c}i{\'c}}, V., {Snedden}, S.~A., {Stebbins}, A., {Steinmetz},
M.,
  {Stoughton}, C., {Strauss}, M.~A., {Subba Rao}, M., {Suto}, Y.,
{Szalay},
  A.~S., {Szapudi}, I., {Szkody}, P., {Tanaka}, M., {Tegmark}, M.,
{Teodoro},
  L.~F.~A., {Thakar}, A.~R., {Tremonti}, C.~A., {Tucker}, D.~L.,
{Uomoto}, A.,
  {Vanden Berk}, D.~E., {Vandenberg}, J., {Vidrih}, S., {Vogeley}, M.~S.,
  {Voges}, W., {Vogt}, N.~P., {Wadadekar}, Y., {Watters}, S., {Weinberg},
  D.~H., {West}, A.~A., {White}, S.~D.~M., {Wilhite}, B.~C., {Wonders},
A.~C.,
  {Yanny}, B., {Yocum}, D.~R., {York}, D.~G., {Zehavi}, I., {Zibetti},
S., \&
  {Zucker}, D.~B. 2009, \apjs, 182, 543

\bibitem[{{Adams} {et~al.}(1988){Adams}, {Shu}, \&
{Lada}}]{adamsetal88-1}
{Adams}, F.~C., {Shu}, F.~H., \& {Lada}, C.~J. 1988, \apj, 326, 865

\bibitem[{{All{\`e}gre} {et~al.}(1995){All{\`e}gre}, {Poirier}, {Humler},
\&
  {Hofmann}}]{allegreetal95-1}
{All{\`e}gre}, C.~J., {Poirier}, J., {Humler}, E., \& {Hofmann}, A.~W.
1995,
  Earth and Planetary Science Letters, 134, 515

\bibitem[{{Bochkarev} \& {Rafikov}(2011)}]{bochkarev+rafikov01-1}
{Bochkarev}, K.~V. \& {Rafikov}, R.~R. 2011, \apj, 741, 36

\bibitem[{{Brinkworth} {et~al.}(2009){Brinkworth}, {G{\"a}nsicke},
{Marsh},
  {Hoard}, \& {Tappert}}]{brinkworthetal09-1}
{Brinkworth}, C.~S., {G{\"a}nsicke}, B.~T., {Marsh}, T.~R., {Hoard},
D.~W., \&
  {Tappert}, C. 2009, \apj, 696, 1402

\bibitem[{{Chiang} \& {Goldreich}(1997)}]{chiang+goldreich97-1}
{Chiang}, E.~I. \& {Goldreich}, P. 1997, \apj, 490, 368

\bibitem[{{Copenhagen University Obs.} {et~al.}(2006){Copenhagen
University
  Obs.}, {Institute of Astronomy, Cambridge}, \& {Real Instituto y
Observatorio
  de la Armada en San Fernando}}]{cmc14}
{Copenhagen University Obs.}, {Institute of Astronomy, Cambridge}, U., \&
{Real
  Instituto y Observatorio de la Armada en San Fernando}. 2006, Carlsberg
  Meridian Catalog Number 14

\bibitem[{{Davidsson}(1999)}]{davidsson99-1}
{Davidsson}, B.~J.~R. 1999, Icarus, 142, 525

\bibitem[{{Debes} {et~al.}(2011){Debes}, {Hoard}, {Kilic}, {Wachter},
  {Leisawitz}, {Cohen}, {Kirkpatrick}, \& {Griffith}}]{debesetal11-1}
{Debes}, J.~H., {Hoard}, D.~W., {Kilic}, M., {Wachter}, S., {Leisawitz},
D.~T.,
  {Cohen}, M., {Kirkpatrick}, J.~D., \& {Griffith}, R.~L. 2011, \apj,
729, 4

\bibitem[{{Debes} \& {Sigurdsson}(2002)}]{debesetal02-1}
{Debes}, J.~H. \& {Sigurdsson}, S. 2002, \apj, 572, 556

\bibitem[{{Desharnais} {et~al.}(2008){Desharnais}, {Wesemael}, {Chayer},
  {Kruk}, \& {Saffer}}]{desharnaisetal08-1}
{Desharnais}, S., {Wesemael}, F., {Chayer}, P., {Kruk}, J.~W., \&
{Saffer},
  R.~A. 2008, \apj, 672, 540

\bibitem[{{Dufour} {et~al.}(2007){Dufour}, {Bergeron}, {Liebert},
{Harris},
  {Knapp}, {Anderson}, {Hall}, {Strauss}, {Collinge}, \&
  {Edwards}}]{dufouretal07-2}
{Dufour}, P., {Bergeron}, P., {Liebert}, J., {Harris}, H.~C., {Knapp},
G.~R.,
  {Anderson}, S.~F., {Hall}, P.~B., {Strauss}, M.~A., {Collinge}, M.~J.,
\&
  {Edwards}, M.~C. 2007, \apj, 663, 1291

\bibitem[{{Dufour} {et~al.}(2010){Dufour}, {Kilic}, {Fontaine},
{Bergeron},
  {Lachapelle}, {Kleinman}, \& {Leggett}}]{dufouretal10-1}
{Dufour}, P., {Kilic}, M., {Fontaine}, G., {Bergeron}, P., {Lachapelle},
F.,
  {Kleinman}, S.~J., \& {Leggett}, S.~K. 2010, \apj, 719, 803

\bibitem[{{Dupuis} {et~al.}(1993){Dupuis}, {Fontaine}, {Pelletier}, \&
  {Wesemael}}]{dupuisetal93-1}
{Dupuis}, J., {Fontaine}, G., {Pelletier}, C., \& {Wesemael}, F. 1993,
\apjs,
  84, 73

\bibitem[{{Eggen}(1968)}]{eggen68-1}
{Eggen}, O.~J. 1968, \apjs, 16, 97

\bibitem[{{Epchtein} {et~al.}(1999){Epchtein}, {Deul}, {Derriere},
  {Borsenberger}, {Egret}, {Simon}, {Alard}, {Bal{\'a}zs}, {de Batz},
{Cioni},
  {Copet}, {Dennefeld}, {Forveille}, {Fouqu{\'e}}, {Garz{\'o}n},
{Habing},
  {Holl}, {Hron}, {Kimeswenger}, {Lacombe}, {Le Bertre}, {Loup}, {Mamon},
  {Omont}, {Paturel}, {Persi}, {Robin}, {Rouan}, {Tiph{\`e}ne},
{Vauglin}, \&
  {Wagner}}]{epchteinetal99-1}
{Epchtein}, N., {Deul}, E., {Derriere}, S., {Borsenberger}, J., {Egret},
D.,
  {Simon}, G., {Alard}, C., {Bal{\'a}zs}, L.~G., {de Batz}, B., {Cioni},
M.-R.,
  {Copet}, E., {Dennefeld}, M., {Forveille}, T., {Fouqu{\'e}}, P.,
  {Garz{\'o}n}, F., {Habing}, H.~J., {Holl}, A., {Hron}, J.,
{Kimeswenger}, S.,
  {Lacombe}, F., {Le Bertre}, T., {Loup}, C., {Mamon}, G.~A., {Omont},
A.,
  {Paturel}, G., {Persi}, P., {Robin}, A.~C., {Rouan}, D., {Tiph{\`e}ne},
D.,
  {Vauglin}, I., \& {Wagner}, S.~J. 1999, \aap, 349, 236

\bibitem[{{Farihi} {et~al.}(2010{\natexlab{a}}){Farihi}, {Barstow},
{Redfield},
  {Dufour}, \& {Hambly}}]{farihietal10-2}
{Farihi}, J., {Barstow}, M.~A., {Redfield}, S., {Dufour}, P., \&
{Hambly},
  N.~C. 2010{\natexlab{a}}, \mnras, 404, 2123

\bibitem[{{Farihi} {et~al.}(2011{\natexlab{a}}){Farihi}, {Brinkworth},
  {G{\"a}nsicke}, {Marsh}, {Girven}, {Hoard}, {Klein}, \&
  {Koester}}]{farihietal11-1}
{Farihi}, J., {Brinkworth}, C.~S., {G{\"a}nsicke}, B.~T., {Marsh}, T.~R.,
  {Girven}, J., {Hoard}, D.~W., {Klein}, B., \& {Koester}, D.
  2011{\natexlab{a}}, \apjl, 728, L8

\bibitem[{{Farihi} {et~al.}(2011{\natexlab{b}}){Farihi}, {G{\"a}nsicke},
  {Steele}, {Girven}, {Burleigh}, {Breedt}, \&
{Koester}}]{farihietal11-3}
{Farihi}, J., {G{\"a}nsicke}, B.~T., {Steele}, P.~R., {Girven}, J.,
{Burleigh},
  M.~R., {Breedt}, E., \& {Koester}, D. 2011{\natexlab{b}}, ArXiv
e-prints

\bibitem[{{Farihi} {et~al.}(2010{\natexlab{b}}){Farihi}, {Jura}, {Lee},
\&
  {Zuckerman}}]{farihietal10-1}
{Farihi}, J., {Jura}, M., {Lee}, J., \& {Zuckerman}, B.
2010{\natexlab{b}},
  \apj, 714, 1386

\bibitem[{{Farihi} {et~al.}(2009){Farihi}, {Jura}, \&
  {Zuckerman}}]{farihietal09-1}
{Farihi}, J., {Jura}, M., \& {Zuckerman}, B. 2009, \apj, 694, 805

\bibitem[{{Farihi} {et~al.}(2008){Farihi}, {Zuckerman}, \&
  {Becklin}}]{farihietal08-1}
{Farihi}, J., {Zuckerman}, B., \& {Becklin}, E.~E. 2008, \apj, 674, 431

\bibitem[{{Fazio} {et~al.}(2004){Fazio}, {Hora}, {Allen}, {Ashby},
{Barmby},
  {Deutsch}, {Huang}, {Kleiner}, {Marengo}, {Megeath}, {Melnick},
{Pahre},
  {Patten}, {Polizotti}, {Smith}, {Taylor}, {Wang}, {Willner},
{Hoffmann},
  {Pipher}, {Forrest}, {McMurty}, {McCreight}, {McKelvey}, {McMurray},
{Koch},
  {Moseley}, {Arendt}, {Mentzell}, {Marx}, {Losch}, {Mayman}, {Eichhorn},
  {Krebs}, {Jhabvala}, {Gezari}, {Fixsen}, {Flores}, {Shakoorzadeh},
{Jungo},
  {Hakun}, {Workman}, {Karpati}, {Kichak}, {Whitley}, {Mann},
{Tollestrup},
  {Eisenhardt}, {Stern}, {Gorjian}, {Bhattacharya}, {Carey}, {Nelson},
  {Glaccum}, {Lacy}, {Lowrance}, {Laine}, {Reach}, {Stauffer}, {Surace},
  {Wilson}, {Wright}, {Hoffman}, {Domingo}, \& {Cohen}}]{fazioetal04-1}
{Fazio}, G.~G., {Hora}, J.~L., {Allen}, L.~E., {Ashby}, M.~L.~N.,
{Barmby}, P.,
  {Deutsch}, L.~K., {Huang}, J., {Kleiner}, S., {Marengo}, M., {Megeath},
  S.~T., {Melnick}, G.~J., {Pahre}, M.~A., {Patten}, B.~M., {Polizotti},
J.,
  {Smith}, H.~A., {Taylor}, R.~S., {Wang}, Z., {Willner}, S.~P.,
{Hoffmann},
  W.~F., {Pipher}, J.~L., {Forrest}, W.~J., {McMurty}, C.~W.,
{McCreight},
  C.~R., {McKelvey}, M.~E., {McMurray}, R.~E., {Koch}, D.~G., {Moseley},
S.~H.,
  {Arendt}, R.~G., {Mentzell}, J.~E., {Marx}, C.~T., {Losch}, P.,
{Mayman}, P.,
  {Eichhorn}, W., {Krebs}, D., {Jhabvala}, M., {Gezari}, D.~Y., {Fixsen},
  D.~J., {Flores}, J., {Shakoorzadeh}, K., {Jungo}, R., {Hakun}, C.,
{Workman},
  L., {Karpati}, G., {Kichak}, R., {Whitley}, R., {Mann}, S.,
{Tollestrup},
  E.~V., {Eisenhardt}, P., {Stern}, D., {Gorjian}, V., {Bhattacharya},
B.,
  {Carey}, S., {Nelson}, B.~O., {Glaccum}, W.~J., {Lacy}, M., {Lowrance},
  P.~J., {Laine}, S., {Reach}, W.~T., {Stauffer}, J.~A., {Surace}, J.~A.,
  {Wilson}, G., {Wright}, E.~L., {Hoffman}, A., {Domingo}, G., \&
{Cohen}, M.
  2004, \apjs, 154, 10

\bibitem[{{Friedrich} {et~al.}(2000){Friedrich}, {Koester}, {Christlieb},
  {Reimers}, \& {Wisotzki}}]{friedrichetal00-1}
{Friedrich}, S., {Koester}, D., {Christlieb}, N., {Reimers}, D., \&
{Wisotzki},
  L. 2000, \aap, 363, 1040

\bibitem[{{G{\"a}nsicke} {et~al.}(2008){G{\"a}nsicke}, {Koester},
{Marsh},
  {Rebassa-Mansergas}, \& {Southworth}}]{gaensickeetal08-1}
{G{\"a}nsicke}, B.~T., {Koester}, D., {Marsh}, T.~R.,
{Rebassa-Mansergas}, A.,
  \& {Southworth}, J. 2008, \mnras, 391, L103

\bibitem[{{G{\"a}nsicke} {et~al.}(2007){G{\"a}nsicke}, {Marsh}, \&
  {Southworth}}]{gaensickeetal07-1}
{G{\"a}nsicke}, B.~T., {Marsh}, T.~R., \& {Southworth}, J. 2007, \mnras,
380,
  L35

\bibitem[{{G{\"a}nsicke} {et~al.}(2006){G{\"a}nsicke}, {Marsh},
{Southworth},
  \& {Rebassa-Mansergas}}]{gaensickeetal06-3}
{G{\"a}nsicke}, B.~T., {Marsh}, T.~R., {Southworth}, J., \&
  {Rebassa-Mansergas}, A. 2006, Science, 314, 1908

\bibitem[{{Girven} {et~al.}(2011){Girven}, {G{\"a}nsicke}, {Steeghs}, \&
  {Koester}}]{girvenetal11-1}
{Girven}, J., {G{\"a}nsicke}, B.~T., {Steeghs}, D., \& {Koester}, D.
2011,
  \mnras, 417, 1210

\bibitem[{{Graham} {et~al.}(1990){Graham}, {Matthews}, {Neugebauer}, \&
  {Soifer}}]{grahametal90-1}
{Graham}, J.~R., {Matthews}, K., {Neugebauer}, G., \& {Soifer}, B.~T.
1990,
  \apj, 357, 216

\bibitem[{{Greenstein}(1956)}]{greenstein56-1}
{Greenstein}, J.~L. 1956, Vistas in Astronomy, 2, 1299

\bibitem[{{Hagen} {et~al.}(1995){Hagen}, {Groote}, {Engels}, \&
  {Reimers}}]{hagenetal95-1}
{Hagen}, H.-J., {Groote}, D., {Engels}, D., \& {Reimers}, D. 1995, \aaps,
111,
  195

\bibitem[{{Houck} {et~al.}(2004){Houck}, {Roellig}, {Van Cleve},
{Forrest},
  {Herter}, {Lawrence}, {Matthews}, {Reitsema}, {Soifer}, {Watson},
{Weedman},
  {Huisjen}, {Troeltzsch}, {Barry}, {Bernard-Salas}, {Blacken}, {Brandl},
  {Charmandaris}, {Devost}, {Gull}, {Hall}, {Henderson}, {Higdon},
{Pirger},
  {Schoenwald}, {Sloan}, {Uchida}, {Appleton}, {Armus}, {Burgdorf},
  {Fajardo-Acosta}, {Grillmair}, {Ingalls}, {Morris}, \&
  {Teplitz}}]{houcketal04-1}
{Houck}, J.~R., {Roellig}, T.~L., {Van Cleve}, J., {Forrest}, W.~J.,
{Herter},
  T.~L., {Lawrence}, C.~R., {Matthews}, K., {Reitsema}, H.~J., {Soifer},
B.~T.,
  {Watson}, D.~M., {Weedman}, D., {Huisjen}, M., {Troeltzsch}, J.~R.,
{Barry},
  D.~J., {Bernard-Salas}, J., {Blacken}, C., {Brandl}, B.~R.,
{Charmandaris},
  V., {Devost}, D., {Gull}, G.~E., {Hall}, P., {Henderson}, C.~P.,
{Higdon},
  S.~J.~U., {Pirger}, B.~E., {Schoenwald}, J., {Sloan}, G.~C., {Uchida},
K.~I.,
  {Appleton}, P.~N., {Armus}, L., {Burgdorf}, M.~J., {Fajardo-Acosta},
S.~B.,
  {Grillmair}, C.~J., {Ingalls}, J.~G., {Morris}, P.~W., \& {Teplitz},
H.~I.
  2004, in Society of Photo-Optical Instrumentation Engineers (SPIE)
Conference
  Series, Vol. 5487, Society of Photo-Optical Instrumentation Engineers
(SPIE)
  Conference Series, ed. {J.~C.~Mather}, 62--76

\bibitem[{{Hunt} {et~al.}(1998){Hunt}, {Mannucci}, {Testi}, {Migliorini},
  {Stanga}, {Baffa}, {Lisi}, \& {Vanzi}}]{huntetal98-1}
{Hunt}, L.~K., {Mannucci}, F., {Testi}, L., {Migliorini}, S., {Stanga},
R.~M.,
  {Baffa}, C., {Lisi}, F., \& {Vanzi}, L. 1998, \aj, 115, 2594

\bibitem[{{Jura}(2003)}]{jura03-1}
{Jura}, M. 2003, \apjl, 584, L91

\bibitem[{{Jura}(2008)}]{jura08-1}
---. 2008, \aj, 135, 1785

\bibitem[{{Jura} {et~al.}(2007){Jura}, {Farihi}, \&
{Zuckerman}}]{juraetal07-1}
{Jura}, M., {Farihi}, J., \& {Zuckerman}, B. 2007, \apj, 663, 1285

\bibitem[{{Jura} {et~al.}(2009){Jura}, {Farihi}, \&
{Zuckerman}}]{juraetal09-1}
---. 2009, \aj, 137, 3191

\bibitem[{{Kilic} {et~al.}(2008){Kilic}, {Farihi}, {Nitta}, \&
  {Leggett}}]{kilicetal08-1}
{Kilic}, M., {Farihi}, J., {Nitta}, A., \& {Leggett}, S.~K. 2008, \aj,
136, 111

\bibitem[{{Kilic} {et~al.}(2006){Kilic}, {von Hippel}, {Leggett}, \&
  {Winget}}]{kilicetal06-1}
{Kilic}, M., {von Hippel}, T., {Leggett}, S.~K., \& {Winget}, D.~E. 2006,
\apj,
  646, 474

\bibitem[{{Klein} {et~al.}(2011){Klein}, {Jura}, {Koester}, \&
  {Zuckerman}}]{kleinetal11-1}
{Klein}, B., {Jura}, M., {Koester}, D., \& {Zuckerman}, B. 2011, \apj,
741, 64

\bibitem[{{Klein} {et~al.}(2010){Klein}, {Jura}, {Koester}, {Zuckerman},
\&
  {Melis}}]{kleinetal10-1}
{Klein}, B., {Jura}, M., {Koester}, D., {Zuckerman}, B., \& {Melis}, C.
2010,
  \apj, 709, 950

\bibitem[{{Koester}(2009)}]{koester09-1}
{Koester}, D. 2009, \aap, 498, 517

\bibitem[{{Koester}(2010)}]{koester10-1}
---. 2010, Memorie della Societa Astronomica Italiana,, 81, 921

\bibitem[{{Koester} {et~al.}(2011){Koester}, {Girven}, {G{\"a}nsicke}, \&
  {Dufour}}]{koesteretal11-1}
{Koester}, D., {Girven}, J., {G{\"a}nsicke}, B.~T., \& {Dufour}, P. 2011,
\aap,
  530, A114+

\bibitem[{{Koester} {et~al.}(2005{\natexlab{a}}){Koester}, {Napiwotzki},
  {Voss}, {Homeier}, \& {Reimers}}]{koesteretal05-1}
{Koester}, D., {Napiwotzki}, R., {Voss}, B., {Homeier}, D., \& {Reimers},
D.
  2005{\natexlab{a}}, \aap, 439, 317

\bibitem[{{Koester} {et~al.}(2005{\natexlab{b}}){Koester}, {Rollenhagen},
  {Napiwotzki}, {Voss}, {Christlieb}, {Homeier}, \&
  {Reimers}}]{koesteretal05-2}
{Koester}, D., {Rollenhagen}, K., {Napiwotzki}, R., {Voss}, B.,
{Christlieb},
  N., {Homeier}, D., \& {Reimers}, D. 2005{\natexlab{b}}, \aap, 432, 1025

\bibitem[{{Koester} \& {Wilken}(2006)}]{koester+wilken06-1}
{Koester}, D. \& {Wilken}, D. 2006, \aap, 453, 1051

\bibitem[{{Lawrence} {et~al.}(2007){Lawrence}, {Warren}, {Almaini},
{Edge},
  {Hambly}, {Jameson}, {Lucas}, {Casali}, {Adamson}, {Dye}, {Emerson},
  {Foucaud}, {Hewett}, {Hirst}, {Hodgkin}, {Irwin}, {Lodieu}, {McMahon},
  {Simpson}, {Smail}, {Mortlock}, \& {Folger}}]{lawrenceetal07-1}
{Lawrence}, A., {Warren}, S.~J., {Almaini}, O., {Edge}, A.~C., {Hambly},
N.~C.,
  {Jameson}, R.~F., {Lucas}, P., {Casali}, M., {Adamson}, A., {Dye}, S.,
  {Emerson}, J.~P., {Foucaud}, S., {Hewett}, P., {Hirst}, P., {Hodgkin},
S.~T.,
  {Irwin}, M.~J., {Lodieu}, N., {McMahon}, R.~G., {Simpson}, C., {Smail},
I.,
  {Mortlock}, D., \& {Folger}, M. 2007, \mnras, 379, 1599

\bibitem[{{Makovoz} {et~al.}(2006){Makovoz}, {Roby}, {Khan}, \&
  {Booth}}]{makovozetal06-1}
{Makovoz}, D., {Roby}, T., {Khan}, I., \& {Booth}, H. 2006, in Society of
  Photo-Optical Instrumentation Engineers (SPIE) Conference Series, Vol.
6274,
  Society of Photo-Optical Instrumentation Engineers (SPIE) Conference
Series,
  10

\bibitem[{{Manchado} {et~al.}(1998){Manchado}, {Fuentes}, {Prada},
  {Ballesteros}, {Barreto}, {Carranza}, {Escudero}, {Fragoso-Lopez},
  {Joven-Alvarez}, {Manescau}, {Pi}, {Rodriguez-Ramos}, \&
  {Sosa}}]{manchadoetal98-1}
{Manchado}, A., {Fuentes}, F.~J., {Prada}, F., {Ballesteros}, E.,
{Barreto},
  M., {Carranza}, J.~M., {Escudero}, I., {Fragoso-Lopez}, A.~B.,
  {Joven-Alvarez}, E., {Manescau}, A., {Pi}, M., {Rodriguez-Ramos},
L.~F., \&
  {Sosa}, N.~A. 1998, in Society of Photo-Optical Instrumentation
Engineers
  (SPIE) Conference Series, Vol. 3354, Society of Photo-Optical
Instrumentation
  Engineers (SPIE) Conference Series, ed. {A.~M.~Fowler}, 448--455

\bibitem[{{Marleau} {et~al.}(2004){Marleau}, {Fadda}, {Storrie-Lombardi},
  {Helou}, {Makovoz}, {Frayer}, {Yan}, {Appleton}, {Armus}, {Chapman},
{Choi},
  {Fang}, {Heinrichsen}, {Im}, {Lacy}, {Shupe}, {Soifer}, {Squires},
{Surace},
  {Teplitz}, \& {Wilson}}]{marleauetal04-1}
{Marleau}, F.~R., {Fadda}, D., {Storrie-Lombardi}, L.~J., {Helou}, G.,
  {Makovoz}, D., {Frayer}, D.~T., {Yan}, L., {Appleton}, P.~N., {Armus},
L.,
  {Chapman}, S., {Choi}, P.~I., {Fang}, F., {Heinrichsen}, I., {Im}, M.,
  {Lacy}, M., {Shupe}, D., {Soifer}, B.~T., {Squires}, G., {Surace}, J.,
  {Teplitz}, H.~I., \& {Wilson}, G. 2004, \apjs, 154, 66

\bibitem[{{Martin} {et~al.}(2005){Martin}, {Fanson}, {Schiminovich},
  {Morrissey}, {Friedman}, {Barlow}, {Conrow}, {Grange}, {Jelinsky},
  {Milliard}, {Siegmund}, {Bianchi}, {Byun}, {Donas}, {Forster},
{Heckman},
  {Lee}, {Madore}, {Malina}, {Neff}, {Rich}, {Small}, {Surber}, {Szalay},
  {Welsh}, \& {Wyder}}]{martinetal05-1}
{Martin}, D.~C., {Fanson}, J., {Schiminovich}, D., {Morrissey}, P.,
{Friedman},
  P.~G., {Barlow}, T.~A., {Conrow}, T., {Grange}, R., {Jelinsky}, P.~N.,
  {Milliard}, B., {Siegmund}, O.~H.~W., {Bianchi}, L., {Byun}, Y.-I.,
{Donas},
  J., {Forster}, K., {Heckman}, T.~M., {Lee}, Y.-W., {Madore}, B.~F.,
{Malina},
  R.~F., {Neff}, S.~G., {Rich}, R.~M., {Small}, T., {Surber}, F.,
{Szalay},
  A.~S., {Welsh}, B., \& {Wyder}, T.~K. 2005, \apjl, 619, L1

\bibitem[{{Melis} {et~al.}(2011){Melis}, {Farihi}, {Dufour}, {Zuckerman},
  {Burgasser}, {Bergeron}, {Bochanski}, \& {Simcoe}}]{melisetal11-1}
{Melis}, C., {Farihi}, J., {Dufour}, P., {Zuckerman}, B., {Burgasser},
A.~J.,
  {Bergeron}, P., {Bochanski}, J., \& {Simcoe}, R. 2011, \apj, 732, 90

\bibitem[{{Moorwood} {et~al.}(1998){Moorwood}, {Cuby}, \&
  {Lidman}}]{moorwoodetal98-1}
{Moorwood}, A., {Cuby}, J.-G., \& {Lidman}, C. 1998, The Messenger, 91, 9

\bibitem[{{Paquette} {et~al.}(1986){Paquette}, {Pelletier}, {Fontaine},
\&
  {Michaud}}]{paquetteetal86-1}
{Paquette}, C., {Pelletier}, C., {Fontaine}, G., \& {Michaud}, G. 1986,
\apjs,
  61, 177

\bibitem[{{Rafikov}(2011{\natexlab{a}})}]{rafikov11-1}
{Rafikov}, R.~R. 2011{\natexlab{a}}, \apjl, 732, L3+

\bibitem[{{Rafikov}(2011{\natexlab{b}})}]{rafikov11-2}
---. 2011{\natexlab{b}}, \mnras, 416, L55

\bibitem[{{Reach} {et~al.}(2005){Reach}, {Kuchner}, {von Hippel},
{Burrows},
  {Mullally}, {Kilic}, \& {Winget}}]{reachetal05-1}
{Reach}, W.~T., {Kuchner}, M.~J., {von Hippel}, T., {Burrows}, A.,
{Mullally},
  F., {Kilic}, M., \& {Winget}, D.~E. 2005, \apjl, 635, L161

\bibitem[{{Reach} {et~al.}(2009){Reach}, {Lisse}, {von Hippel}, \&
  {Mullally}}]{reachetal09-1}
{Reach}, W.~T., {Lisse}, C., {von Hippel}, T., \& {Mullally}, F. 2009,
\apj,
  693, 697

\bibitem[{{Sion} {et~al.}(1990){Sion}, {Leckenby}, \&
{Szkody}}]{sionetal90-1}
{Sion}, E.~M., {Leckenby}, H.~J., \& {Szkody}, P. 1990, \apjl, 364, L41

\bibitem[{{Skrutskie} {et~al.}(2006){Skrutskie}, {Cutri}, {Stiening},
  {Weinberg}, {Schneider}, {Carpenter}, {Beichman}, {Capps}, {Chester},
  {Elias}, {Huchra}, {Liebert}, {Lonsdale}, {Monet}, {Price}, {Seitzer},
  {Jarrett}, {Kirkpatrick}, {Gizis}, {Howard}, {Evans}, {Fowler},
{Fullmer},
  {Hurt}, {Light}, {Kopan}, {Marsh}, {McCallon}, {Tam}, {Van Dyk}, \&
  {Wheelock}}]{skrutskieetal06-1}
{Skrutskie}, M.~F., {Cutri}, R.~M., {Stiening}, R., {Weinberg}, M.~D.,
  {Schneider}, S., {Carpenter}, J.~M., {Beichman}, C., {Capps}, R.,
{Chester},
  T., {Elias}, J., {Huchra}, J., {Liebert}, J., {Lonsdale}, C., {Monet},
D.~G.,
  {Price}, S., {Seitzer}, P., {Jarrett}, T., {Kirkpatrick}, J.~D.,
{Gizis},
  J.~E., {Howard}, E., {Evans}, T., {Fowler}, J., {Fullmer}, L., {Hurt},
R.,
  {Light}, R., {Kopan}, E.~L., {Marsh}, K.~A., {McCallon}, H.~L., {Tam},
R.,
  {Van Dyk}, S., \& {Wheelock}, S. 2006, \aj, 131, 1163

\bibitem[{{Steele} {et~al.}(2011){Steele}, {Burleigh}, {Dobbie},
{Jameson},
  {Barstow}, \& {Satterthwaite}}]{steeleetal11-1}
{Steele}, P.~R., {Burleigh}, M.~R., {Dobbie}, P.~D., {Jameson}, R.~F.,
  {Barstow}, M.~A., \& {Satterthwaite}, R.~P. 2011, \mnras, 416, 2768

\bibitem[{{Subasavage} {et~al.}(2007){Subasavage}, {Henry}, {Bergeron},
  {Dufour}, {Hambly}, \& {Beaulieu}}]{subasavageetal07-1}
{Subasavage}, J.~P., {Henry}, T.~J., {Bergeron}, P., {Dufour}, P.,
{Hambly},
  N.~C., \& {Beaulieu}, T.~D. 2007, \aj, 134, 252

\bibitem[{{van Maanen}(1917)}]{vanmaanen17-1}
{van Maanen}, A. 1917, \pasp, 29, 258

\bibitem[{{Vennes} {et~al.}(2010){Vennes}, {Kawka}, \&
  {N{\'e}meth}}]{vennesetal10-1}
{Vennes}, S., {Kawka}, A., \& {N{\'e}meth}, P. 2010, \mnras, 404, L40

\bibitem[{{Vennes} {et~al.}(2011){Vennes}, {Kawka}, \&
  {N{\'e}meth}}]{vennesetal11-1}
---. 2011, \mnras, 413, 2545

\bibitem[{{von Hippel} {et~al.}(2007){von Hippel}, {Kuchner}, {Kilic},
  {Mullally}, \& {Reach}}]{vonhippeletal07-1}
{von Hippel}, T., {Kuchner}, M.~J., {Kilic}, M., {Mullally}, F., \&
{Reach},
  W.~T. 2007, \apj, 662, 544

\bibitem[{{Voss} {et~al.}(2007){Voss}, {Koester}, {Napiwotzki},
{Christlieb},
  \& {Reimers}}]{vossetal07-1}
{Voss}, B., {Koester}, D., {Napiwotzki}, R., {Christlieb}, N., \&
{Reimers}, D.
  2007, \aap, 470, 1079

\bibitem[{{Weidemann}(1960)}]{weidemann60-1}
{Weidemann}, V. 1960, \apj, 131, 638

\bibitem[{{Wisotzki} {et~al.}(1996){Wisotzki}, {Koehler}, {Groote}, \&
  {Reimers}}]{wisotzkietal96-1}
{Wisotzki}, L., {Koehler}, T., {Groote}, D., \& {Reimers}, D. 1996,
\aaps, 115,
  227

\bibitem[{{Wright} {et~al.}(2010){Wright}, {Eisenhardt}, {Mainzer},
{Ressler},
  {Cutri}, {Jarrett}, {Kirkpatrick}, {Padgett}, {McMillan}, {Skrutskie},
  {Stanford}, {Cohen}, {Walker}, {Mather}, {Leisawitz}, {Gautier},
{McLean},
  {Benford}, {Lonsdale}, {Blain}, {Mendez}, {Irace}, {Duval}, {Liu},
{Royer},
  {Heinrichsen}, {Howard}, {Shannon}, {Kendall}, {Walsh}, {Larsen},
{Cardon},
  {Schick}, {Schwalm}, {Abid}, {Fabinsky}, {Naes}, \&
{Tsai}}]{wrightetal10-1}
{Wright}, E.~L., {Eisenhardt}, P.~R.~M., {Mainzer}, A.~K., {Ressler},
M.~E.,
  {Cutri}, R.~M., {Jarrett}, T., {Kirkpatrick}, J.~D., {Padgett}, D.,
  {McMillan}, R.~S., {Skrutskie}, M., {Stanford}, S.~A., {Cohen}, M.,
{Walker},
  R.~G., {Mather}, J.~C., {Leisawitz}, D., {Gautier}, T.~N., {McLean},
I.,
  {Benford}, D., {Lonsdale}, C.~J., {Blain}, A., {Mendez}, B., {Irace},
W.~R.,
  {Duval}, V., {Liu}, F., {Royer}, D., {Heinrichsen}, I., {Howard}, J.,
  {Shannon}, M., {Kendall}, M., {Walsh}, A.~L., {Larsen}, M., {Cardon},
J.~G.,
  {Schick}, S., {Schwalm}, M., {Abid}, M., {Fabinsky}, B., {Naes}, L., \&
  {Tsai}, C. 2010, \aj, 140, 1868

\bibitem[{{Xu} \& {Jura}(2012)}]{xu+jura12-1}
{Xu}, S. \& {Jura}, M. 2012, \apj, 745, 88

\bibitem[{{Zuckerman} \& {Becklin}(1987)}]{zuckerman+becklin87-1}
{Zuckerman}, B. \& {Becklin}, E.~E. 1987, \nat, 330, 138

\bibitem[{{Zuckerman} {et~al.}(2011){Zuckerman}, {Koester}, {Dufour},
{Melis},
  {Klein}, \& {Jura}}]{zuckermanetal11-1}
{Zuckerman}, B., {Koester}, D., {Dufour}, P., {Melis}, C., {Klein}, B.,
\&
  {Jura}, M. 2011, \apj, 739, 101

\bibitem[{{Zuckerman} {et~al.}(2007){Zuckerman}, {Koester}, {Melis},
{Hansen},
  \& {Jura}}]{zuckermanetal07-1}
{Zuckerman}, B., {Koester}, D., {Melis}, C., {Hansen}, B.~M., \& {Jura},
M.
  2007, \apj, 671, 872

\bibitem[{{Zuckerman} {et~al.}(2003){Zuckerman}, {Koester}, {Reid}, \&
  {H{\"u}nsch}}]{zuckermanetal03-1}
{Zuckerman}, B., {Koester}, D., {Reid}, I.~N., \& {H{\"u}nsch}, M. 2003,
\apj,
  596, 477

\bibitem[{{Zuckerman} {et~al.}(2010){Zuckerman}, {Melis}, {Klein},
{Koester},
  \& {Jura}}]{zuckermanetal10-1}
{Zuckerman}, B., {Melis}, C., {Klein}, B., {Koester}, D., \& {Jura}, M.
2010,
  \apj, 722, 725

\end{thebibliography}

\begin{deluxetable}{cccccccrcccc}
\tabletypesize{\scriptsize}
\tablecaption{\textit{Spitzer} IRAC White Dwarf Targets\label{t-tar}}
\tablewidth{0pt}
\tablehead{
\colhead{WD}							&
\colhead{Name}						&
\colhead{SpT}							&
\colhead{$V$}							&
\colhead{$M$}							&
\colhead{$T_{\rm eff}$}					&
\colhead{$t_{\rm cool}$}					&
\colhead{[H/He]}						&
\colhead{[Ca/H(e)]}						&
\colhead{log $\langle dM_{\rm z}/dt \rangle$}	&
\colhead{log ($M_{\rm z}$)}				&
\colhead{Ref}			\\
\colhead{}							&
\colhead{}							&
\colhead{}							&
\colhead{(mag)}					&
\colhead{($M_{\odot}$)}				&
\colhead{(K)}						&
\colhead{(Gyr)}						&
\colhead{}							&
\colhead{}							&
\colhead{(g\,s$^{-1}$)}				&
\colhead{(g)}						&
\colhead{}}

\startdata

\textit{0110$-$565}			&HE\,0110$-$5630	&DBAZ	&15.9	&0.71	&19\,200	&0.12	&$-4.2$	&$-7.9$				&8.4		&20.2   	&1,2\\
\textit{0435$+$410}			&GD\,61 			&DBAZ	&14.9	&0.72	&17\,300	&0.20	&$-4.0$	&$-7.5$				&8.9		&21.5   	&3\\
0446$-$255				&HE\,0446$-$2531	&DBAZ	&16.8	&0.59	&12\,600	&0.36	&$-3.2$	&$-5.7$\tablenotemark{b}	&11.5   	&24.9   	&4\\
0449$-$259				&HE\,0449$-$2554	&DBAZ	&16.3	&0.72	&12\,500	&0.52	&$-4.5$	&$-7.2$\tablenotemark{b}	&10.0	&23.5   	&4\\
0802$+$386				&G111-54			&DZ		&15.5	&0.78	&11\,000	&0.85	&$<-6.0$	&$-9.8$				&7.4		&21.1   	&5\\
0838$+$375				&CBS\,78			&DBZ	&17.7	&0.59	&12\,500	&0.36	&$<-5.0$	&$-8.0$				&9.2		&22.7   	&6\\
0953$+$594				&SDSS   			&DZA	&18.4	&0.58	&8\,200 	&1.12	&$-3.2$	&$-7.5$				&9.7		&23.6   	&5\\
1015$+$377				&CBS\,127   		&DZ		&17.7	&0.58	&10\,500	&0.59	&$<-5.0$	&$-8.2$				&9.1		&22.8   	&6\\
\textit{1349$-$230}			&HE\,1349$-$2305	&DBAZ	&16.5	&0.67	&18\,200	&0.14	&$-4.7$	&$-8.0$				&8.7		&22.0   	&1,2\\
1350$-$162				&HE\,1350$-$1612	&DBAZ	&17.0	&0.77	&15\,000	&0.35	&$-4.7$	&$-6.7$\tablenotemark{b}	&10.4   	&23.4   	&4\\
1352$+$004				&PG  			&DBAZ	&15.8	&0.59	&15\,300	&0.20	&$-4.8$	&$-9.3$				&7.7		&20.7   	&1\\
1614$+$160				&PG 				&DAZ	&15.6	&0.50	&17\,400	&0.10	&\nodata	&$-7.2$				&7.8		&\nodata	&1\\
2138$-$332				&NLTT\,51844		&DZ		&14.5	&0.69	&7\,200 	&2.23	&\nodata	&$-8.6$				&8.5		&22.5   	&7\\
2142$-$169\tablenotemark{a}
						&PHL\,131   		&DO		&15.8	&\nodata	&\nodata	&\nodata	&\nodata	&\nodata				&\nodata  	&\nodata	&2\\
2229$+$139				&PG  			&DBAZ	&16.0	&0.83	&15\,900	&0.36	&$-4.5$	&$-9.0$				&8.0		&20.9   	&1\\
2322$+$118				&LB\,1188   		&DZA	&16.0	&0.59	&12\,000	&0.41	&$-5.2$	&$-8.7$				&8.5		&22.1   	&1\\

\enddata

\tablenotetext{a}{Not metal-rich.}

\tablenotetext{b}{\citet{kleinetal11-1} find that HS\,2253$+$8023 has a
lower calcium abundance by around 1.0\,dex than originally reported. This
may therefore also be true of other \citet{friedrichetal00-1} stars.}

\tablecomments{Objects found to have dust disks are listed in italics. 
The tenth column lists time-averaged metal in-fall rates. The eleventh
column shows the mass of metals in the convective envelopes of the
helium-rich stars. Both quantities are calculated based on the observed
calcium abundances and assuming this represents 1.6\,per\,cent of the
total mass of heavy elements, as in the bulk Earth
\cite{allegreetal95-1}.}

\tablerefs{
(1) \citealt{koesteretal05-1};
(2) \citealt{vossetal07-1};
(3) \citealt{desharnaisetal08-1};
(4) \citealt{friedrichetal00-1};
(5) \citealt{dufouretal07-2};
(6) \citealt{dupuisetal93-1};
(7) \citealt{subasavageetal07-1}}

\end{deluxetable}

\clearpage

\begin{deluxetable}{ccc}
\tabletypesize{\scriptsize}
\tablecaption{IRAC Coordinates for HE\,and HS\,White Dwarfs \label{t-hshe}}
\tablewidth{0pt}
\tablehead{
\colhead{Star}				&
\colhead{$\alpha$}			&
\colhead{$\delta$}}

\startdata

HE\,0110$-$5630		&$01^{\rm h} 12^{\rm m} 21.15^{\rm s}$	&$-56\arcdeg 14' 27\farcs8$\\
HE\,0446$-$2531		&$04^{\rm h} 49^{\rm m} 01.39^{\rm s}$	&$-25\arcdeg 26' 36\farcs1$\\
HE\,0449$-$2554		&$04^{\rm h} 51^{\rm m} 53.72^{\rm s}$	&$-25\arcdeg 49' 14\farcs7$\\
HE\,1349$-$2305		&$13^{\rm h} 52^{\rm m} 44.13^{\rm s}$	&$-23\arcdeg 20' 05\farcs4$\\
HE\,1350$-$1612		&$13^{\rm h} 53^{\rm m} 34.96^{\rm s}$	&$-16\arcdeg 27' 06\farcs6$\\

\enddata

\tablecomments{Epoch 2009 positions as measured on the IRAC array from image header astrometry.}

\end{deluxetable}

\begin{deluxetable}{ccccc}
\tabletypesize{\scriptsize}
\tablecaption{\textit{Spitzer} IRAC and IRS fluxes.\label{t-flu}}
\tablewidth{0pt}
\tablehead{
\colhead{WD}					&
\colhead{$F_{3.6\mu{\rm m}}$}		&
\colhead{$F_{4.5\mu{\rm m}}$}		&
\colhead{$F_{7.9\mu{\rm m}}$}		&
\colhead{$F_{15.6\mu{\rm m}}$}\\
\colhead{}						&
\colhead{($\mu$Jy)}				&
\colhead{($\mu$Jy)}				&
\colhead{($\mu$Jy)}				&
\colhead{($\mu$Jy)}
}             

\startdata
0110$-$565 				&$135\pm9$ 	&$124\pm8$  	&\nodata          			&\nodata\\
0435$+$410 				&\nodata         	&$259\pm13$ 	&$153\pm8$  			&$98\pm10$\\
0446$-$255 				&\nodata         	&$26\pm1$   	&$15$\tablenotemark{b} &\nodata\\
0449$-$259 				&\nodata         	&$39\pm2$   	&$24\pm1$   			&\nodata\\
0802$+$386				&\nodata         	&$85\pm4$   	&$38\pm2$   			&\nodata\\
0838$+$375 				&$18\pm1$  	&$13\pm1$   	&\nodata          			&\nodata\\
0953$+$594 				&\nodata         	&$9\pm1$    	&$6$\tablenotemark{b}  	&\nodata\\
1015$+$377 				&$20\pm1$  	&$14\pm1$   	&\nodata          			&\nodata\\
1349$-$230 				&\nodata         	&$88\pm2$   	&$66\pm3$   			&\nodata\\
1350$-$162 				&\nodata         	&$23\pm2$   	&$7$\tablenotemark{b} 	&\nodata\\
          					&$32\pm2$  	&$23\pm1$   	&\nodata          			&\nodata\\
1352$+$004 				&\nodata         	&$52\pm3$   	&$15\pm2$   			&\nodata\\
1614$+$160 				&\nodata         	&$63\pm3$   	&$32\pm2$   			&\nodata\\
2138$-$332 				&\nodata         	&$503\pm25$ 	&$195\pm10$ 			&$95\pm15$\\
2142$-$169\tablenotemark{a}
 						&\nodata    	&$35\pm2$   	&$12\pm4$   			&\nodata\\
2229$+$139 				&\nodata        	&$43\pm2$   	&$13\pm2$   			&\nodata\\
2322$+$118 				&$105\pm5$ 	&$69\pm3$   	&\nodata          			&\nodata\\

\enddata

\tablenotetext{a}{Not metal-polluted.}

\tablenotetext{b}{$3\sigma$ upper limit.}

\tablecomments{A 5\,per\,cent uncertainty has been folded in to account
for calibration errors \citep{reachetal05-1}. Pipeline S18.7.0 was used to
obtain BCD products for all targets.}

\end{deluxetable}

\clearpage

\begin{deluxetable}{lcccl}
\tabletypesize{\scriptsize}
\tablecaption{Near-Infrared Photometry\label{t-ind}}
\tablewidth{0pt}
\tablehead{
\colhead{WD}				&
\colhead{$J$}                 	&
\colhead{$H$}                 	&
\colhead{$K$}                 	&
\colhead{Instrument}\\
\colhead{}				&
\colhead{(mag)}                 	&
\colhead{(mag)}                 	&
\colhead{(mag)}                 	&
\colhead{}
}                

\startdata
0110$-$565 	&$16.23\pm0.05$ 	&$16.27\pm0.05$ 	&$16.24\pm0.05$ 	&SOFI  \\
0435$+$410 	&$15.24\pm0.05$ 	&$15.16\pm0.05$ 	&$15.12\pm0.05$ 	&LIRIS \\
0446$-$255 	&$17.06\pm0.05$ 	&$17.09\pm0.05$ 	&$17.15\pm0.05$ 	&SOFI  \\
0449$-$259 	&$16.52\pm0.05$ 	&$16.56\pm0.05$ 	&$16.60\pm0.05$ 	&SOFI  \\
0802$+$386 	&$15.60\pm0.05$ 	&$15.59\pm0.05$ 	&$15.65\pm0.05$ 	&LIRIS \\
0838$+$375 	&$18.03\pm0.05$ 	&$18.02\pm0.05$ 	&$17.98\pm0.05$ 	&LIRIS \\
0953$+$594 	&$18.33\pm0.06$ 	&$18.29\pm0.06$ 	&$18.23\pm0.06$ 	&WFCAM\tablenotemark{a}\\
1015$+$377 	&$17.24\pm0.10$ 	&$17.31\pm0.12$ 	&$16.78\pm0.31$ 	&GEMINI\tablenotemark{b}\\
1349$-$230 	&$16.91\pm0.05$ 	&$16.94\pm0.05$ 	&$16.78\pm0.05$ 	&SOFI  \\
1350$-$162 	&$17.13\pm0.05$ 	&$17.17\pm0.05$ 	&$17.25\pm0.05$ 	&SOFI  \\
1352$+$004 	&$16.10\pm0.05$ 	&$16.12\pm0.05$ 	&$16.15\pm0.05$ 	&LIRIS \\
1614$+$160 	&$15.93\pm0.05$ 	&$15.97\pm0.05$ 	&$16.04\pm0.05$ 	&SOFI  \\
2229$+$139 	&$16.29\pm0.05$ 	&$16.36\pm0.05$ 	&$16.53\pm0.05$ 	&SOFI  \\
2322$+$118 	&$15.99\pm0.05$ 	&$15.98\pm0.05$ 	&$16.03\pm0.05$ 	&SOFI  \\

\enddata

\tablenotetext{a}{Obtained by one of us at UKIRT using WFCAM (Casali et al.\ 2007, A\&A, 467, 777).}
\tablenotetext{b}{Obtained by C. Melis at Lick Observatory with the GEMINI camera (McLean et al.\ 1993, 
SPIE, 1946, 513).}

\end{deluxetable}

\clearpage

\end{document}